\documentclass[11pt]{article}
\usepackage[pdftex]{graphicx}
\begin{document}

\begin{center}
\textbf{\Large A Triginometric Seasonal Component Model \\[2mm] and its Application to Time Series
with Two Types of Seasonality}

\vspace{7mm}
{\large Genshiro Kitagawa\\[3mm]

The Institute of Statistical Mathematics\\
and\\
Graduate University for Advanced Study
}

\end{center}

\vspace{5mm}

\begin{center}{\bf\large Abstract}\end{center}

\begin{quote}
A finite trigonometric series model for seasonal time series is considered in this paper.
This component model is shown to be useful, in particular, for the modeling of time series with
two types of seasonality, a long-period and a short period.
This component model is also shown to be effective in the case of ordinary seasonal time series with only one seasonal component, if the seasonal pattern is simple and can be well represented by a small number of trigonometric components.
As examples, electricity demand data, bi-hourly temperature data, CO${}_2$ data, and two economic
time series are considered.
The last section summarizes the findings from the emperical studies.
\end{quote}

\vspace{5mm}

\section{Introduction}\label{introduction}
This paper is addressed to the problem of modeling time series with seasonal variation 
based on trigonometric series component model.
Various model based methods have been proposed which explicitly use statistical models 
(Cleveland and Tiao (1976), Box et al. (1978), Akaike (1979), Akaike and lshiguro (1982), 
Hillmer and Tiao (1982), Gersch and Kitagawa (1983), and Taylor, J. W. (2010)). 

In this paper, we report our experience in replacing the seasonal component model used in the Decomp model (Kitagawa and Gersch (1984,1996) and Kitagawa(2021)) with a trigonometric model. In particular, we report on the case of two seasonal components with different periods, such as daily and weekly or daily and yearly.
When two seasonal component models of the decomposition type are used simultaneously, it is necessary to modify the model to guarantee the uniqueness of the decomposition, but in the case of the trigonometric model, it is found that it is sufficient to eliminate the components with overlapping frequencies from the periodic component model with a long period.

In the standard Decomp model, the time series with seasonality is expressed as
\begin{eqnarray}
y_n &=& T_n + S_n + p_n + w_n,
\end{eqnarray}
where $y_n$ is the time series and $T_n$, $S_n$, $p_n$ and $w_n$ are
the trend, the seasonal component, the stationary AR component and
the observation noise, respectively.
These components are assumed to follow the component models
\begin{eqnarray}
\Delta^{m_1}T_n &=& v_n^{(T)} \nonumber \\
\left( \sum_{j=0}^{p-1} B^j \right)^{m_2}S_n &=&  v_n^{(S)} \label{eq_component models} \\ 
p_n &=& \sum_{j=1}^{m_3} a_j p_{n-j} + v_n^{(p)},
.\nonumber
\end{eqnarray}
where $B$ is the back-shift operator, $B x_n \equiv x_{n-1}$ and $Delta = 1-B$.

The basic model and the component models can be expressed in
a state-space model form
\begin{eqnarray}
x_n &=& Fx_{n-1} + Gv_n \nonumber \\
y_n &=& Hx_n + w_n,
\end{eqnarray}
where, for example, for $m_1=2$ and $m_2=1$, the state and the system noise are defined by 
$x_n =(T_n, T_{n-1},S_n,$ $S_{n-1},\ldots , S_{n-p+1},p_n,\ldots ,p_{n-m_3+1})^T$
and $v_n =(v_n^{(t)},v_n^{(s)},v_n^{(p)})$, and
$F$, $G$ and $H$ are defined by
{\setlength\arraycolsep{1mm}
\begin{eqnarray}
F &=& \left[ \begin{array}{cc|cccc|ccc}\arraycolsep=0.1mm
             2 & -1&   &   &   &   &   &   &   \\
             1 & 0 &   &   &   &   &   &   &   \\ \hline
               &   & -1& -1&\cdots&-1& &   &   \\
               &   & 1 &   &   &   &   &   &   \\
               &   &   &\ddots&&   &   &   &   \\
               &   &   &   & 1 &   &   &   &   \\ \hline
               &   &   &   &   &   &a_1&\cdots&a_{m_3} \\
               &   &   &   &   &   & 1 &   &\\
               &   &   &   &   &   &   & 1 &   
       \end{array}\right],\quad
G = \left[ \begin{array}{c|c|c}
             1 &   &   \\
             0 &   &   \\ \hline
               & 1 &   \\
               & 0 &   \\
               &\vdots&   \\
               & 0 &   \\ \hline
               &   & 1 \\
               &   &\vdots\\
               &   & 0    
       \end{array}\right], \\
H &=& \left[ \begin{array}{cc|cccc|ccc} 
                    1 & 0 & 1 & 0 & \cdots & 0 & 1 & \cdots & 0 \end{array}\right].
\end{eqnarray}
}
It is also assumed that the system noise $v_n$ and the observation noise $w_n$
are distributed according to a Gaussian white noises
$v_n\sim N(0,Q)$ and $w_n \sim N(0,\sigma^2 )$, respectively.

Then, given a time series, $y_1,\ldots ,y_N$, the smoothed estimates of the components, 
$t_{n|N}$, $S_{n|N}$ and $p_{n|N}$ are obtained by recursive filtering and smoothing methods.
The parameters of the model is obtained by the maximum likelihood method and the
goodness of the model is evaluated by the AIC.
The log-likelihood and the AIC are defined by 
\begin{eqnarray} 
\ell (\theta ) &=& -\frac{1}{2}\left\{ N\log 2\pi + \sum_{n=1}^N \log r_n
  +\sum_{n=1}^N \frac{\varepsilon_n^2}{r_n}\right\} \nonumber \\
{\rm AIC} &=& -2\ell (\hat{\theta}) + 2(\mbox{number of parameters of the model}),
\end{eqnarray}
where $\varepsilon_n$ and $r_n$ are one-step-ahead prediction error and 
its variance obtained by
\begin{eqnarray}
\varepsilon_n &=& y_n - H x_{n|n-1} \nonumber \\
r_n &=& \sigma^2 + H V_{n|n-1}H^T.
\end{eqnarray}
The number of parameters of the model is given by
$id(m_1)+id(m_2)+id(m_3)+1+ \textrm{dim}(x_n)$ where the function $id(m)$ is deifined by 
\begin{eqnarray}
id(m) = \left\{ \begin{array}{ll} 1 & {\rm if }\:\: m>0 \\
                                  0 & {\rm if }\:\: m=0 \end{array}\right.
\end{eqnarray}
and dim$(x_n$) denotes the dimension of the state vector $x_n$.

In this paper, we consider the use of finite trigonometric series model shown in the
next section 
for the seasonal component instead of the one used in (\ref{eq_component models}).
This is motivated by the anlysis of long temperature record which has both 
daily and annual cycles, in which the trigonometric series model will be 
used as the annual seasonal component.
It is obvious that in this case it is impractical to use the seasonal component
model (\ref{eq_component models}), since the state dimension for the annual
seasonal component becomes $24\times 3654$ for the hourly data.
However, it will be shown that this component model is also useful in the
ordinary seasonal adjustment problem.

The plan of the paper is as follows.
The trigonometric series type seasonal component and the associated state-space
representation is shown in section 2.
Section 3 consider the ordinary seasonal adjustment based on the trigonometric
seasonal model. The model is compared with the conventional Decomp model
for three data sets, CO${}_2$ data, Blsallfood data and the Whard data.
Section 4 presents a  seasonal adjustment model with two types of seasonality, a long period and a short period and as an example, we consider the modeling of almosr four-year temperature data measured every 2 hours.
Concluding remark is given in section 5.

\section{Trigonometric Seasonal Adjustment Model for Seasonality}\index{trigonometric seasonal adjustment model} 

In this section, we briefly summarize the trigonometric-function-based seasonal adjustment method.

\subsection{Trigonometric Component Model for Seasonality}\index{trigonometric component model for seasonality}  

Trigonometric seasonal component model with period $p$ is defined by
%
\begin{eqnarray}
Q_n = \sum_{j=1}^{k_c} c_{jn} \cos (\omega jn) + \sum_{j=1}^{k_s} d_{jn} \sin (\omega jn) ,\quad
\omega =\frac{2\pi}{p},\quad k_c, k_s \leq \frac{p}{2},
\end{eqnarray}
where $\sin (\omega jn)=0$ for $j=p/2$. 
The function of a particular period can be expressed as a weighted sum of Sin and cos, so $k_c=k_s$ usually. However, when $k_s=\frac{p}{2}$, $\sin (\omega k_s n)$=0, so the maximum order $\sin$ sunction is invalid. Therefore, the maximum number of terms in the trigonometric model with period $p$ is $p-1$.

\subsection{State-space representation of seasonal component models}
The trigonometric component model is expressed in state-space form as
\begin{eqnarray}
x_n &=& F_n x_{n-1} + G_nv_n \nonumber \\
Q_n &=& H_n x_n, 
\label{eq_SSM_with_AR_noise}
\end{eqnarray}
where $x_n$ is $k_c+k_s$-dimensional state vector,
$v_n$ is a $k_c+k_s$-dimensional state noise.
Here, $v_n$ is a white noise with mean vector 0 and the variance covariance matrix
$\Sigma_n$.

%
%
For example, if $k_c=k_s=k$, $x_n$, $F$, $G$ and $H_n$ are given respectively by
\begin{eqnarray}
x_n &=&\left[ \begin{array}{c}
        c_1 \\ d_1 \\ \vdots \\ c_k \\ d_k \end{array}\right], \quad
F = G = \left[ \begin{array}{ccccc}
                         1&  &  &  &  \\
                          & 1&  &  &  \\
                          &  &\ddots& &\\
                          &  &  & 1 &  \\
                          &  &  &  & 1 \end{array}\right],\\
H_n&=&\left[ \begin{array}{ccccc} 
          \cos(\omega n), & \sin (\omega n), & \cdots ,& \cos(\omega kn), & \sin (\omega kn)\end{array}\right].
\end{eqnarray}

\subsection{Seasonal Adjustment Model}
We consider a model for time series with seasonality,
\begin{eqnarray}
y_n = t_n + p_n + Q_n + w_n ,
\end{eqnarray}
where $t_n$ is the trend component, $p_n$ is the stationary AR process,
$Q_n$ is the trigonometric seasonal component and $w_n$ is the observation noise component.
The period of the seasonality is assumed to be $p$.
Assuming $m_1=2$, these components follow
\begin{eqnarray}
t_n &=& 2t_{n-1} - t_{n-2} + v_n^{(t)} \nonumber \\
p_n &=& \sum_{j=1}^{m_3} a_j p_{n-j}  + v_n^{(p)} \\
Q_n &=& \sum_{j=1}^{k_c} c_{jn} \cos (\omega jn) + \sum_{j=1}^{k_s} d_{jn} \sin (\omega jn). \nonumber
\end{eqnarray}
If the length of the season is $p$, then $\omega =\frac{2\pi}{p}$ and $k_c, k_s \leq \frac{p}{2}$. 
However, since $\sin (\omega \frac{p}{2n})$=0, $k_s$ is less than $\frac{p}{2}$, and the maximum order is $p-1$.

For specific example, $m_1=2$, $m_3=3$ and $m_4=k$, the basic observation model for the time series $y_n$ and the component models are 
expressed in a state-space model as follows
\begin{eqnarray}
 x_n\hspace*{-2mm} &=&\hspace*{-2mm} \left[ \begin{array}{c}
          t_n \\ t_{n-1} \\ \hline p_n \\ p_{n-1}\\ \hline
          c_1 \\ d_1 \\ \vdots \\ c_k \\ d_k \end{array}\right], \quad
\arraycolsep=1mm
F = \left[ \begin{array}{cc|cc|ccccc}
        2 &-1 &   &   &  &  &  &  &  \\
        1 & 0 &   &   &  &  &  &  &  \\ \hline
          &   &a_1&a_2&  &  &  &  &  \\
          &   & 1 & 0 &  &  &  &  &  \\ \hline
          &   &   &   & 1&  &  &  &  \\
          &   &   &   &  & 1&  &  &  \\
          &   &   &  &   &  &\ddots &  &  \\
          &   &   &  &   &  &  & 1&  \\
          &   &   &  &   &  &  &  & 1\end{array}\right],\quad
G = \left[ \begin{array}{c|c|ccccc}
        1 &  &  &  &  & & \\
          &  &  &  &  & &\\ \hline
          & 1&  &  &  & &\\
          &  &  &  &  & &\\ \hline
          &  & 1&  &  & &\\
          &  &  & 1&  & &\\
          &  &  &  & \ddots& &\\
          &  &  &  &  & 1& \\
          &  &  &  &  &  & 1\\
 \end{array}\right], \nonumber \\
H_n\hspace*{-2mm}&=&\hspace*{-2mm}\left[ \begin{array}{ccccccccccc} 
        1 & 0 & 1 &0 & \sin(\omega n) & \cos (\omega n) & \cdots & \sin(\omega kn) & \cos (\omega kn)\end{array}\right].
\end{eqnarray}

\section{Seasonal Adjustment with Single Seasonality}
In this section, using CO${}_2$ data and economic time series, we examine a trigonometric seasonal adjustment model, comparing it with a Decomp-type model for the case of a single periodic component.

\subsection{Example:  CO2 data}

\begin{figure}[tbp]
\begin{center}
\includegraphics[width=75mm,angle=0,clip=]{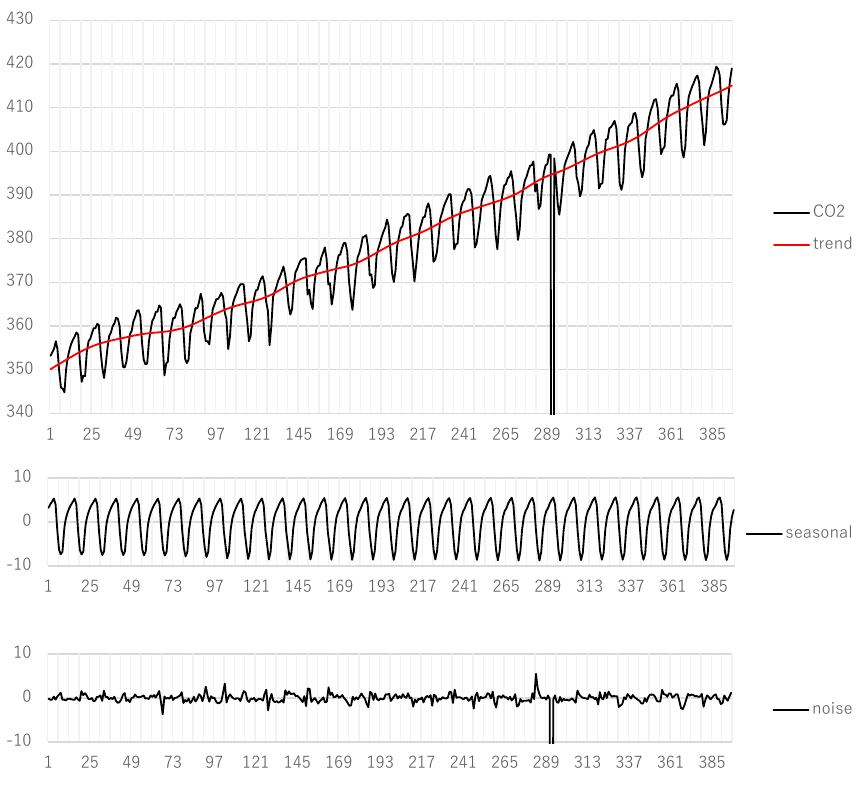}
\includegraphics[width=75mm,angle=0,clip=]{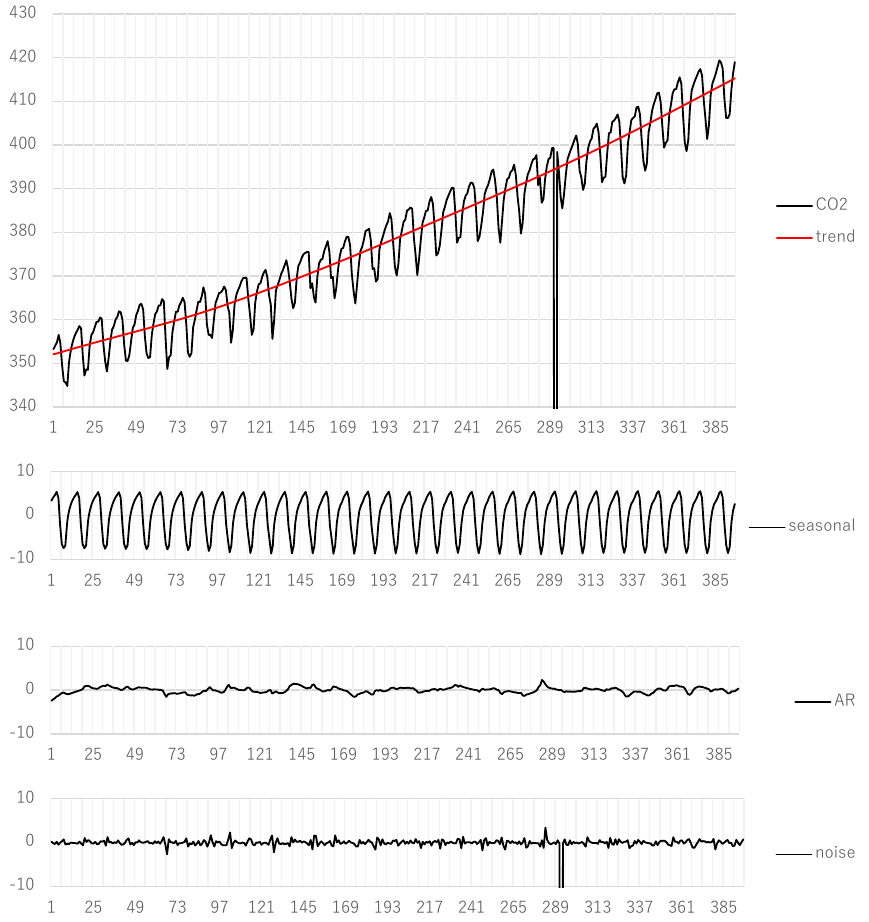}
\end{center}
\caption{Decomposition of CO${}_2$ data by Decomp. Left: $m_1=2$, $m_2=1$, $m_3=0$, right: $m_1=2$, $m_2=1$, $m_3=2$. }\label{Fig_decomp_CO2}
\end{figure}

We consider here the monthly average of carbon dioxide concentration from January 1987 to December 2023 observed at Ayasato, Ohfunato city, Iwate, Japan, released by the Japan Meteorological Agency.
This data contains a missing observation at $n$=292 (April 2011) due to the Tohoku earthquake.
Figure \ref{Fig_decomp_CO2} shows the results by the DECOMP. The left plots show the CO${}_2$ data and estimated trend, seasonal components and noise component by the standard settings, i.e., 
$m_1=2$, $m_2=1$, $m_3=0$.
On the other hand, the right plots show the decompsition by the model with stationary AR components, i.e., $m_1=2$, $m_2=1$, $m_3=1$.
Accoring to the AIC's shown in the bottom line of Table \ref{Tab_AIC_CO2}, the model with AR component ($m_3$=1) is much better than the standard model (i.e., $m_3=0$).

\newpage

Table \ref{Tab_AIC_CO2} also shows the log-likelihoods and the AIC's
of the trigonometric seasonal adjustement models with $m_3=0$, 1, 2 and $m_4=0, 1, \ldots , 11$.
It can be seen that for all of $m_3$=0, 1 and 2, the seasonal models with $m_4=7$ attain the minimum of AIC's
and is considered to be better than the DECOMP model.
This is probably because the seasonal component of the CO${}_2$ data
is rather smooth and only about a half of the sinusoidal components are
necessary to express the seasonal component.

\begin{figure}[tbp]
\begin{center}
\includegraphics[width=80mm,angle=0,clip=]{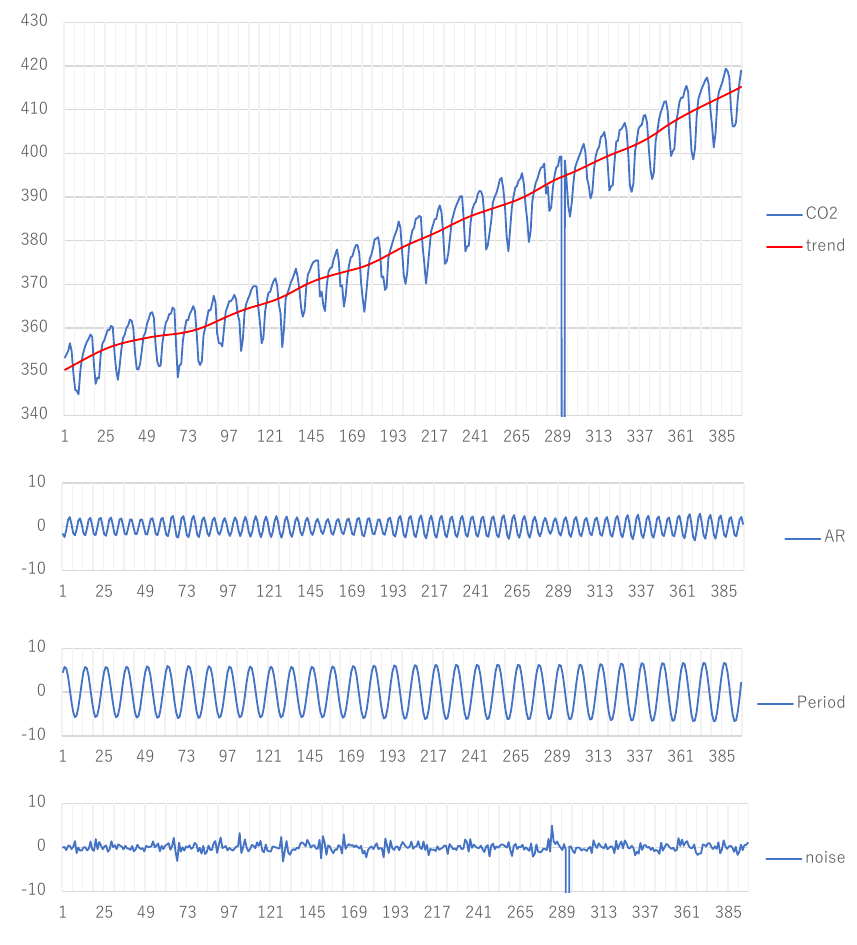}
\includegraphics[width=80mm,angle=0,clip=]{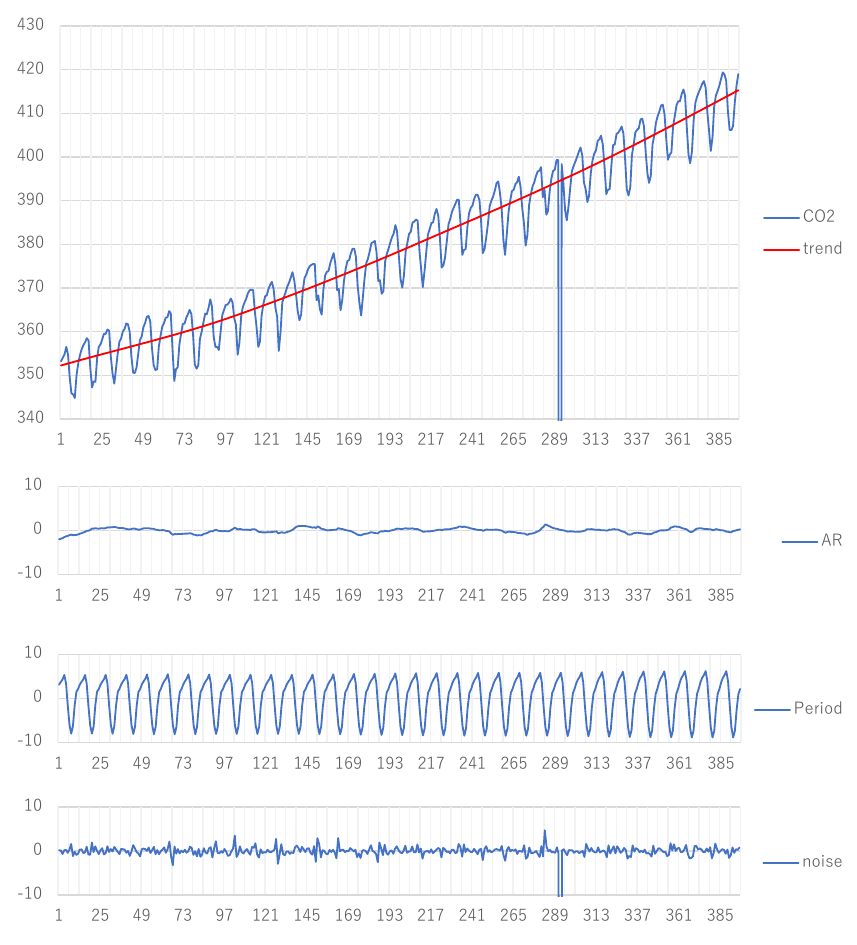}
\end{center}
\caption{Decomposition of CO${}_2$ data by trigonometric seasonal model. 
Left: $m_1=2$, $m_3=2$, $m_4=4$, 
right: $m_1=2$, $m_3=2$, $m_4=6$. }\label{Fig_new_model_CO2}
\end{figure}

Figure \ref{Fig_new_model_CO2} shows the decomposition of the CO${}_2$ data 
by the trigonometric seasonal adjustement models.
The seasonal pattern obtained with $m_4=4$ on the left side does not fully capture the variation pattern of the data, while the result with $m_4=7$ on the right side reproduces the seasonal pattern of the data well.

\begin{table}[bp]
\begin{center}
\caption{The log-likelihood and AIC of the models with various orders for CO${}_2$ data}
\label{Tab_AIC_CO2}

\vspace{2mm}
\begin{tabular}{c|cc|cc|cc}
\hline
         & \multicolumn{2}{c|}{$m_3=0$} & \multicolumn{2}{c|}{$m_3=1$} & \multicolumn{2}{c}{$m_3=2$} \\
   $m_4$ & log-likelihood & {\rm AIC} & log-likelihood & {\rm AIC}& log-likelihood & {\rm AIC} \\
\hline
 0  &-1350.73& 2705.47 & -1138.51 & 2285.02 & -1005.42 & 2020.84 \\
 1  &-1330.47& 2666.74 & -1123.50 & 2256.99 & -1002.32 & 2016.65 \\
 2  &-985.46 & 1976.91 &  -941.11 & 1892.23 & -983.08 & 1978.16 \\
 3  &-978.86 & 1963.72 &  -976.41 & 1962.82 & -976.41 & 1964.83 \\
 4  &-757.01 & 1520.01  & -744.54 & 1499.09 & -744.49 & 1506.97 \\
 5  &-758.77 & 1523.55  & -746.05 & 1502.10 & -745.65 & 1503.30 \\
 6  &-762.57 & 1531.14  & -749.89 & 1509.77 & -749.49 & 1510.98 \\
 7  &-750.64 & \textbf{1507.29}  & -734.75 & \textbf{1479.50} & -734.64 & \textbf{1481.29} \\
 8  &-754.60 & 1531.19  & -738.80 & 1487.61 & -738.68 & 1489.39 \\
 9  &-752.78 & 1529.57  & -735.64 & 1481.28 & -735.61 & 1483.21 \\
10  &-754.55 & 1535.09  & -736.88 & 1483.77 & -736.88 & 1485.75 \\
11  &-758.82 & 1545.64  & -741.27 & 1492.54 & -741.27 & 1494.54 \\
\hline
Decomp& -780.78 & 1567.57 &  -759.39 & \textbf{1528.79} & -759.41 & 1530.83 \\
\hline\end{tabular}
\end{center}
\end{table}

\begin{table}[tbp]
\begin{center}
\caption{The log-likelihood and AIC of the trigonometric regression models for seasonal component}
\label{Tab_AIC_regression}

\vspace{2mm}
\begin{tabular}{c|crr|rrr}
\hline
ORDER &    SIG2     &    AIC   &  AIC-AICMIN &  SIG2     &    AIC   & \# of models\\
\hline
   0&  22.46536  &  2643.734 &   2851.771& 22.46536 & 2643.734 &   1 \\
   1&  20.1321*  &  2597.045 &   2805.081&  5.21549 & 1997.342 &  11 \\
   2&   2.88219  &  1736.014 &   1944.050&  2.65957 & 1700.322 &  55 \\
   3&   2.69533  &  1708.253 &   1916.289&  0.32673 &  770.728 & 165 \\
   4&   1.39415  &   395.203 &    603.239&  0.13941 &  395.203 & 330 \\
   5&   1.26701  &   354.748 &    562.784&  0.04878 &  -69.095 & 462 \\
   6&   1.25989  &   354.243 &    562.279&  0.03606 & -201.173 & 462 \\
   7&   0.35350  &  -208.036 &      0.000&  0.03535 & -208.036 & 330 \\
   8&   0.35350  &  -206.041 &      1.995&  0.03535 & -206.050 & 165 \\
   9&   0.35348  &  -204.055 &      3.982&  0.03535 & -204.056 &  55 \\
  10&   0.35348  &  -202.056 &      5.981&  0.03535 & -202.061 &  11 \\
  11&   0.35348  &  -200.062 &      7.975&  0.03535 & -200.062 &   1 \\
\hline\end{tabular}
\end{center}
\end{table}

\begin{figure}[tbp]
\begin{center}
\includegraphics[width=80mm,angle=0,clip=]{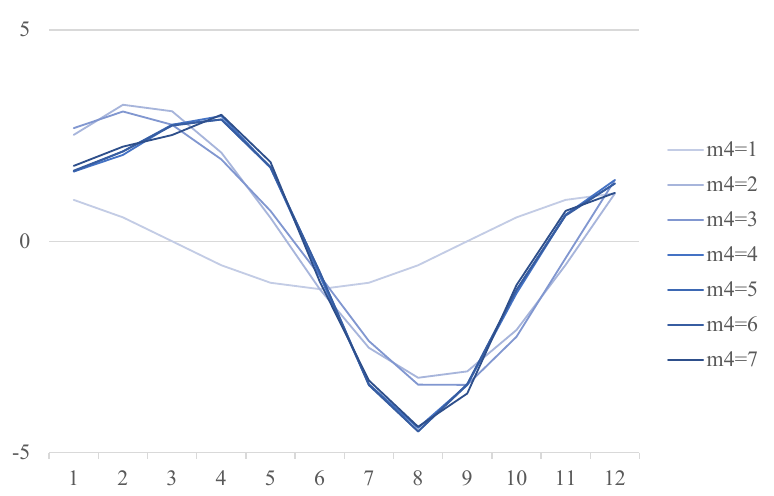}
\end{center}
\caption{Seasonal patterns for Ayasato co${}_2$ data. $m_4=1,\ldots ,7$.}\label{Fig_seasonal_partter_of_CO2_data}
\end{figure}

The left half of Table 2 shows the results of fitting a trigonometric model to the seasonal components estimated by Decomp; the AIC is minimized at order 7, indicating that it is not necessary to use all 11 coefficients.
Regression model coefficients of the 11th order are as follows:
$ a$ =(21.602, 5.8736, 0.61132, -2.2609, -0.15946,  0.037757, -0.42577,
-0.00089,  0.001450, -0.000391,  0.0006976 ).
Since the explanatory variables in the trigonometric model are orthogonal, the coefficients of the seventh-order model are the same as the first seven coefficients of the full-order model.
Figure \ref{Fig_seasonal_partter_of_CO2_data} shows the seasonal patterns estimated by the trigonometric models with orders 1$\sim$7. The eighth-order and higher models are visually indistinguishable from the seventh-order model.

The right half of the table shows the results of selecting the necessary variables using a subset regression model, which also has order 7, and the variables are 1, 2, 3, 4, 5, 6, and 7, which is the same as the simple regression model.

\newpage
\subsection{Example: Blsallhood data}

\begin{figure}[tbp]
\begin{center}
\includegraphics[width=160mm,angle=0,clip=]{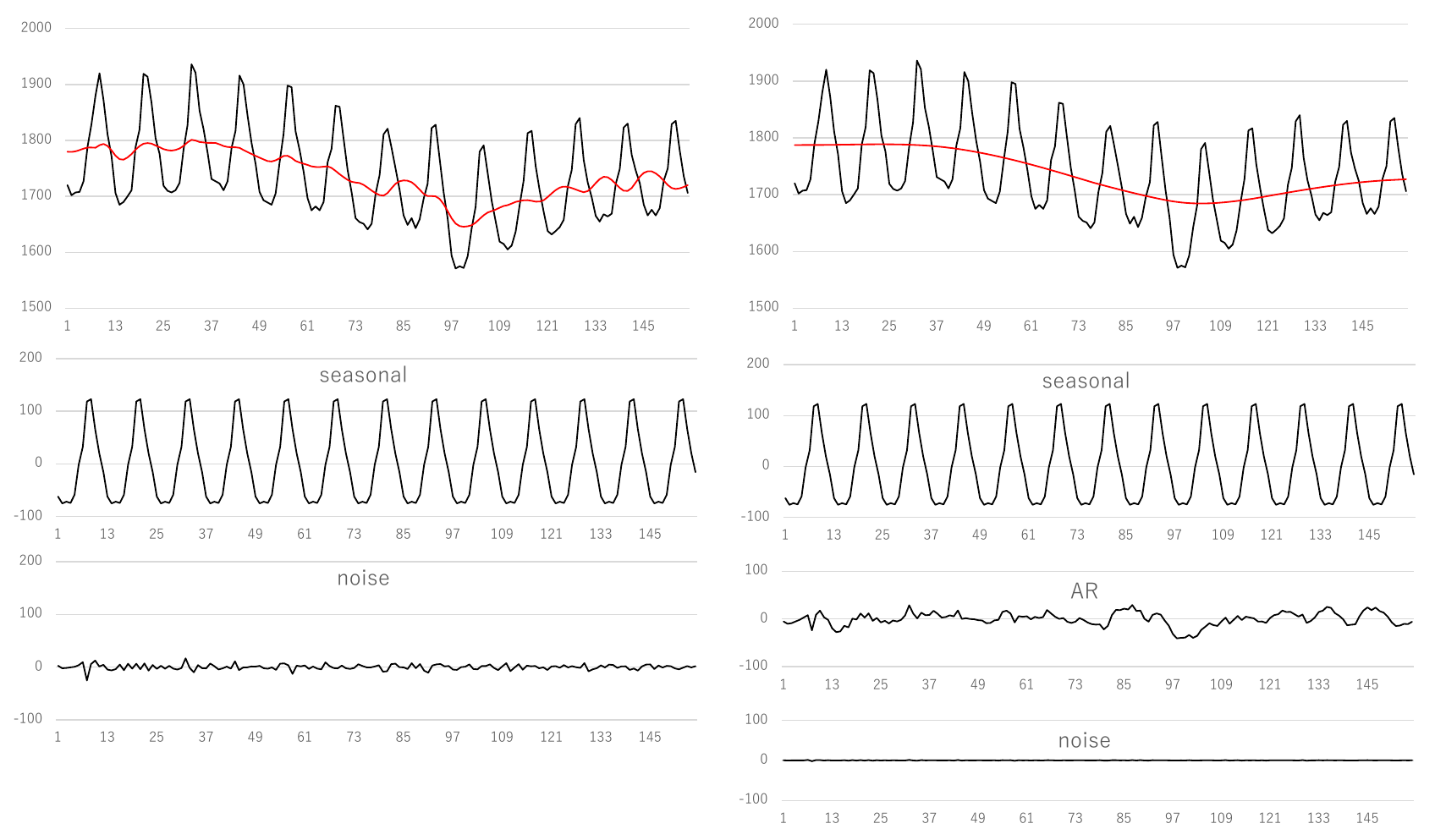}
\includegraphics[width=160mm,angle=0,clip=]{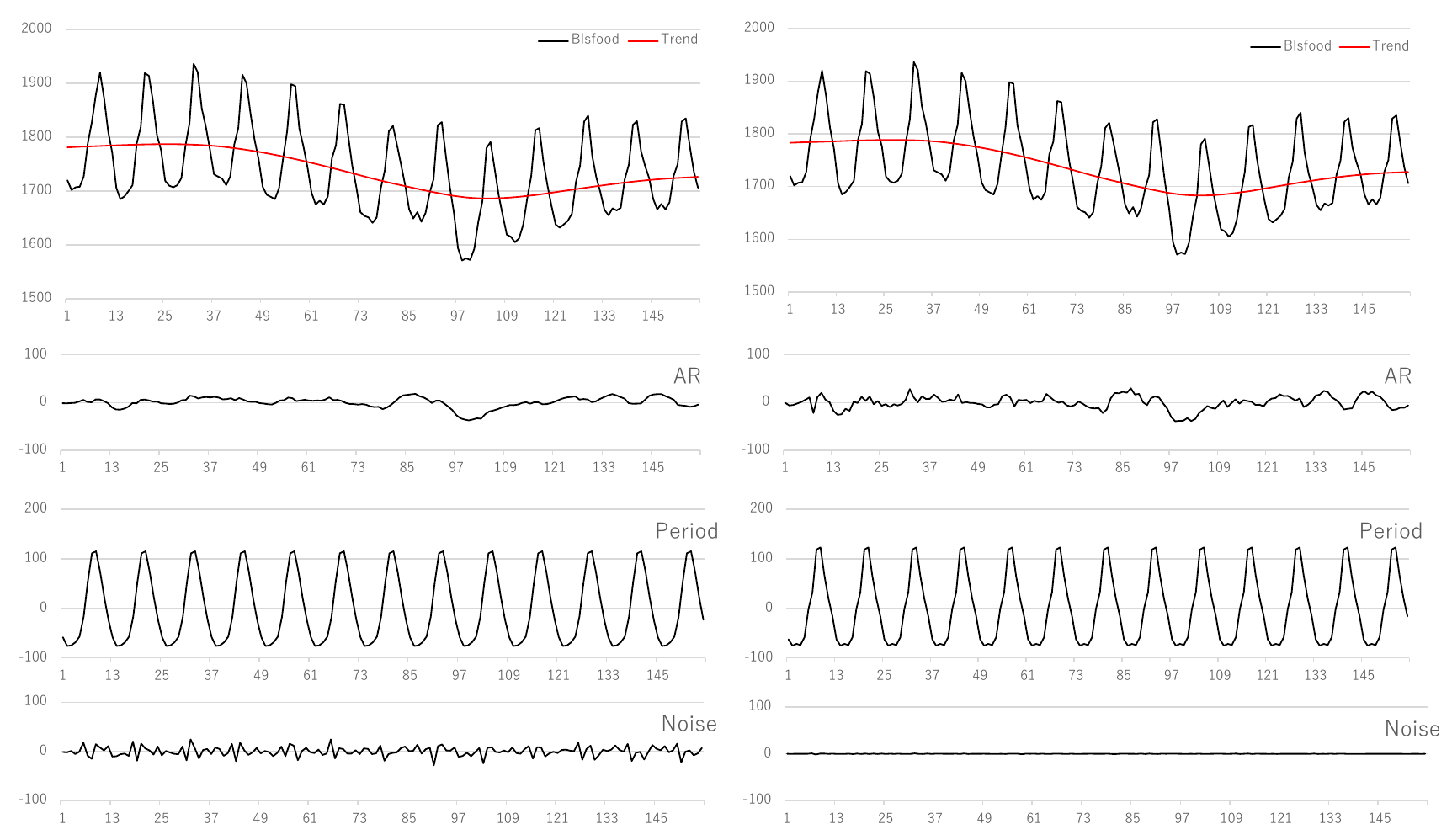}
\caption{Upper plots: Decomposition by DECOMP. Left: $m_1=2$, $m_2=1$, $m_3=0$, 
Right: $m_1=2$, $m_2=1$, $m_3=2$. 
Bottom plots: Decomposition by new seasonal model. Left: $m_1=2$, $m_3=2$, $m_4=8$, 
Right: $m_1=2$, $m_3=2$, $m_4=11$. }\label{Fig_blsfood_decomp}
\end{center}
\end{figure}

Blsallhood data is the number of employees in food industries from January 1967 to December 1979 released by the US Bureau of Labor Statistics. 
Upper plots of Figure \ref{Fig_blsfood_decomp} show the results by the DECOMP. The left plots show the Blsallfood data, estimated trend, seasonal components and noise component by the standard settings, i.e., 
$m_1=2$, $m_2=1$, $m_3=0$.
On the other hand, the right plots show the decomposition by the model with stationary AR components, i.e., $m_1=2$, $m_2=1$, $m_3=2$.
Accoring to the AIC shown in the bottom line of Table \ref{Tab_blsfood_AIC}, the model with AR component is 
considerably smaller than the standard model. 
The estimated trend by the model with $m_3=1$ is significantly better than the one 
with the standard seasonal adjustement model.


\begin{table}[htbp]
\begin{center}
\caption{The log-likelihood and AIC of the models with various orders for Blsallhood data}
\label{Tab_blsfood_AIC}

\vspace{2mm}
\begin{tabular}{c|cc|cc|cc}
\hline
         &\multicolumn{2}{c|}{$m_3=0$} &\multicolumn{2}{c|}{$m_3=1$} &\multicolumn{2}{c}{$m_3=2$} \\
   $m_4$ & log-likelihood & {\rm AIC} &log-likelihood & {\rm AIC} & log-likelihood & {\rm AIC} \\
\hline
   0 & -900.25 & 1806.51 &-808.64 & 1625.27 &-751.04 & 1512.08 \\
   1 & -906.24 & 1820.49 &-814.38 & 1638.75 &-754.11 & 1520.22 \\
   2 & -739.57 & 1489.14 &-724.58 & 1461.15 &-708.79 & 1431.58 \\
   3 & -703.21 & 1418.41 &-696.67 & 1407.33 &-696.67 & 1409.33 \\
   4 & -694.57 & 1403.14 &-688.23 & 1392.47 &-688.25 & 1394.51 \\
   5 & -689.62 & 1395.25 &-682.46 & 1382.93 &-682.36 & 1384.73 \\
   6 & -690.16 & 1398.32 &-682.66 & 1385.32 &-682.45 & 1386.89 \\
   7 & -675.23 & 1370.46 &-664.10 & 1350.21 &-662.55 & 1349.11 \\
   8 & -663.99 & 1349.98 &-650.55 & 1325.09 &-646.25 & 1318.50 \\
   9 & -657.10 & 1338.20 &-642.10 & 1310.20 &-640.03 & 1308.06 \\
  10 & -659.76 & 1345.53 &-644.66 & 1317.32 &-642.75 & 1315.50 \\
  11 & -650.68 & \textbf{1329.37} &-631.98 & \textbf{1293.96} &-632.01 & \textbf{1296.01} \\
\hline
DECOMP & -671.76 & 1349.53 & -653.81 & \textbf{1317.63} & -653.80 & 1319.61 \\
\hline \end{tabular}
\end{center}
\end{table}

Table \ref{Tab_blsfood_AIC} also shows the log-likelihoods and the AIC's
of the trigonometric seasonal adjustement models with $m_3=1,2,3$ and $m_4=0, 1, \ldots , 11$.
In this case, the highest order model, i.e., the one with $m_3=1$ and $m_4=11$ attains the minimum of AIC's
and is considered to be better than the DECOMP model.
Different from the CO${}_2$ data, AIC of the model with highest order $m_4=11$ is significantly smaller than those of the smaller order models. 
Actually, the estimated seasonal component by the model with small order $m_4$ cannot precisely represent the seasonal pattern seen in the Blsallfood data.

Lower panels of Figure \ref{Fig_blsfood_decomp} show the decomposition by the trigonometric
seasonal adjustment model with $m_4=8$ (left panels) and $m_4=11$ (right panels), respectively.
The decomposition with full-order model ($m_4=11$) is quite similar to the ones by the Decomp
model shown in the upper right panels.
On the other hand, seasonal pattern obtained with $m_4=8$ looks too simple.

\newpage
\subsection{Example: Whard data}

\begin{figure}[tbp]
\begin{center}
\includegraphics[width=160mm,angle=0,clip=]{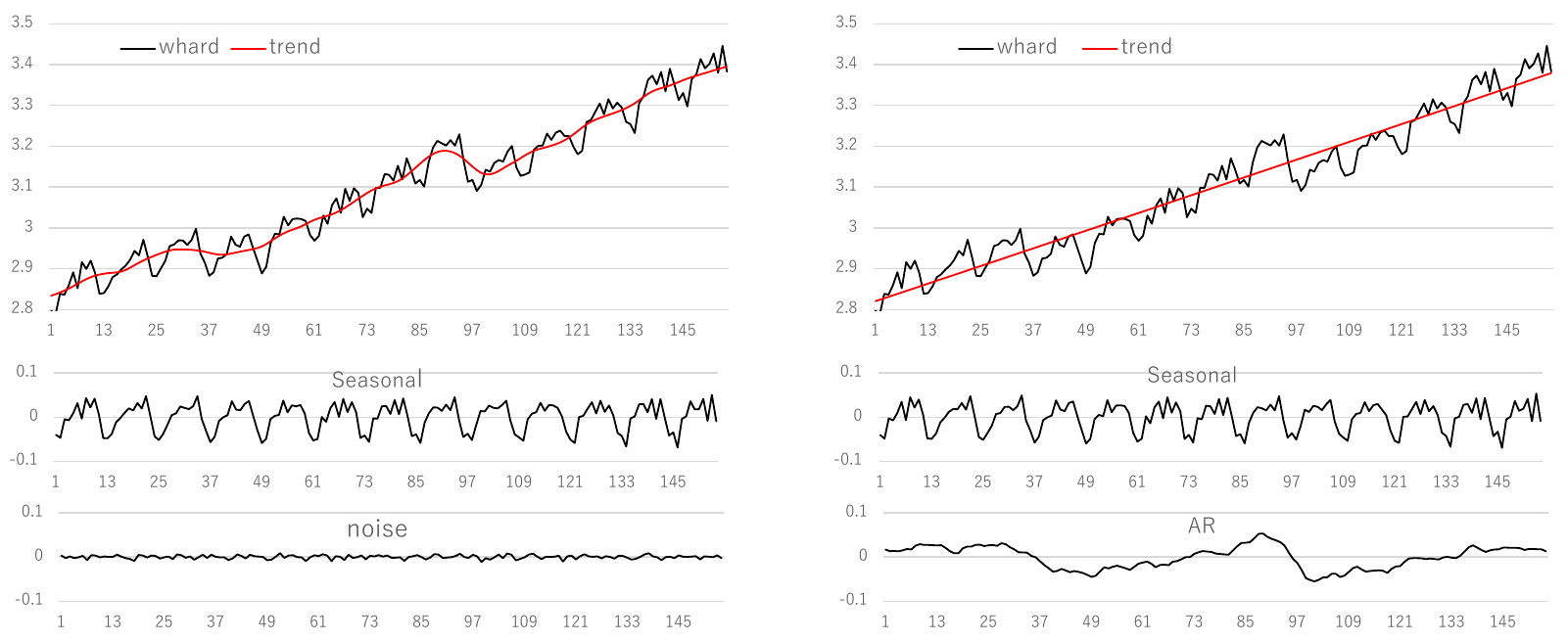}
\includegraphics[width=160mm,angle=0,clip=]{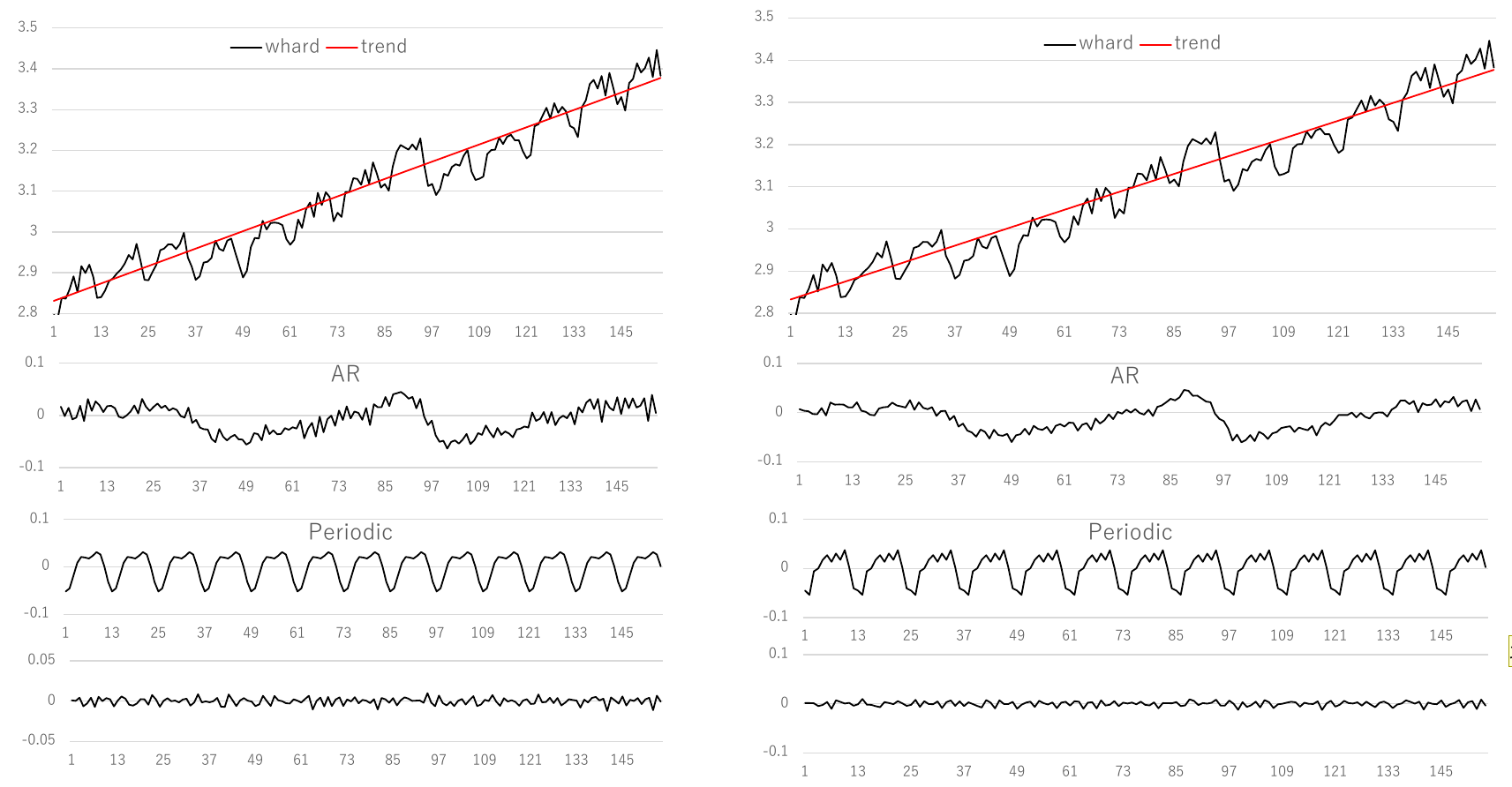}
\end{center}
\caption{Seasonal adjustment of Whard data. Upper plots: Decomposition by DECOMP. Left: $m_1=2$, $m_2=1$, $m_3=0$, Right: $m_1=2$, $m_2=1$, $m_3=2$. Lower plots: Decomposition by new seasonal model. Left: $m_1=2$, $m_3=2$, $m_4=6$, 
Right: $m_1=2$, $m_3=2$, $m_4=11$. }\label{Fig_whard_decomp}
\end{figure}

Whard data shows sales of wholesale hardware companies from January 1967 to November 1979 released by the US Bureau of Labor Statistics. 
Upper panels of Figure \ref{Fig_whard_decomp} shows the results by the DECOMP. The left plots show the Whard data, estimated trend, seasonal components and noise component by the standard settings, i.e., 
$m_1=2$, $m_2=1$, $m_3=0$.
On the other hand, the right plots show the decompsition by the model with stationary AR components, i.e., $m_1=2$, $m_2=1$, $m_3=2$.
Accoring to the AIC shown in Table \ref{Tab_AIC_whard}, the model with AR component is considerably smaller than the standard model. 
The estimated trend by the standard seasonal adjsutment model is wiggly. 
On the other hand, the one by the model with $m_3=2$ is almost a straight line.
It is known that the model with the stationary AR component has better 
ability of long term prediction (Kitagawa 2020).


\begin{table}[htbp]
\begin{center}
\caption{The log-likelihood and AIC of the models with various orders for Whard data}
\index{Tab_AIC_whard}

\vspace{2mm}
\begin{tabular}{cccc}
model   & $m_4$ & log-likelihood & {\rm AIC} \\
\hline
   &  0 & 320.86 & -631.72 \\
   &  2 & 323.17 & -632.35 \\
   &  4 & 345.38 & -672.76 \\
Trigonometric   &  6 & 378.66 & -735.33 \\
   &  8 & 380.69 & -735.38 \\
   & 10 & 377.16 & -724.32 \\
   & 11 & 392.48 & \textbf{-750.97} \\
\hline
Decomp &$m_3=0$& 360.06 & -713.55 \\
       &$m_3=2$& 366.57 & \textbf{-724.81} \\
\hline \end{tabular}\label{Tab_AIC_whard}
\end{center}
\end{table}

Table \ref{Tab_AIC_whard} also shows the log-likelihoods and the AIC's
of the trigonometric seasonal adjustement models with $m_4=0, 2, \ldots , 11$.
In this case, the model with $m_4=6$ attains the minimum of AIC's
but is considerably larger than the DECOMP models with both 
$m_3=2$ and $m_3=2$. 
This is probably because that the seasonal pattern of the Whard data
has a high frequency component that fluctuates every other time.
This means that for this data, it is difficult to express the seasonal compornent
with a small number of sinusoidal components.

Lower panels of Figure \ref{Fig_whard_decomp} show the decomposition by the trigonometric
seasonal adjustment models. The right panels show the results by the AIC best
full-order model ($m_4=11$). In this case, the decomposition is relatively similar to the 
one by the Decomp model with $m_3=2$. However, the estimated seasonal component
is stable than the one by the Decomp model.
The left panels show the results by them model with $m_4=6$.
In this case, the seasonal pattern is too smooth and cannot capture the characteristics
of the Whard data.

\newpage
\section{Two Trigonometric Components Model}

Here we consider seasonal adjustment of time series with two different periods by the following model:
\begin{eqnarray}
y_n &=& T_n + p_n + Q_n^{(1)} +Q_n^{(2)} + w_n , 
\end{eqnarray}
where $y_n$, $T_n$ and $w_n$ are the original series, trend component and noise components as in the previous section. 
$Q_n^{(1)}$ and $Q_n^{(2)}$ are periodic components with periods $f_1$ and $f_2$, respectively, defined by
\begin{eqnarray}
 Q_n^{(1)} &=& \sum_{j=1}^k c_{j,n} \cos (\omega_1 jn) + \sum_{j=1}^k d_{j,n} \sin (\omega_1 jn) \nonumber \\
 Q_n^{(2)} &=& \sum_{j=1}^k c_{j,n} \cos (\omega_2 jn) + \sum_{j=1}^k d_{j,n} \sin (\omega_2 jn),
\end{eqnarray}
where $\omega_1=\frac{2\pi}{f_1}$, $\omega_2=\frac{2\pi}{f_2}$.
It is assmed that $f_1 < f_2$.
If $f_2$ is an integer multiple of $f_1$ and can be expressed as $f_2 = k f_1$, then to avoid multicollinearity, we constrain $j$ in $Q_n(2)$ to not contain multiples of $k$.

\subsection{Hourly Electricity Consumption Data}
As an example, consider logarithm of the Tokyo Electric Power Company's hourly electricity consumption data ($N=2568$,  January 15 - Asugust 30, 2024).
Since the hourly data for electricity is considered to contain a daily cycle and a weekly cycle, we set $f_1$ = 24 and $f_2 = 24\times 7$ = 168. 
Therefore, the maximum order of the model for components $Q_1(n)$ and $Q_2(n)$ is 23 and 167, respectively. 
For simplicity, $Q_1(n)$ is assumed to be full-order, i.e., $m_4$=23. 
Although the order of $Q_2(n)$ can be up to 167, only small orders are actually considered since the objective this component is to capture the rough variation over a week. 
Further, as mentioned above, cycles that are multiples of 7 are included in $Q_1(n)$, so the components $\cos(\omega_2 jn)$ and $\sin(\omega_2jn)$, $j=7,14,\ldots$ are assumed not to be included in $Q_2(n)$.
Table \ref{Tab_AIC_Toden_data} and Figure \ref{Fig_Toden_AIC} show the values of AIC when the number of parameters $m_5$ of $Q_2(n)$ is increased from 0 to 23.
The AIC is minimum at $m_5$=16.

\begin{table}[bp]
\begin{minipage}{0.60\textwidth}
\begin{center}
\caption{The AIC of the two-periodic models with various orders of $m_5$ for Toden electricity consumption data. $m_4=23$}
\label{Tab_AIC_Toden_data}
\begin{tabular}{cc|cc|cc}
 $m_5$ & AIC & $m_5$ & AIC & $m_5$ & AIC \\
\hline
   0 & -18370.47 & 8 & -18513.18 & 16 & \textbf{-18671.97} \\
   1 & -18516.06 & 9 & -18517.87 & 17 & -18666.00 \\
   2 & -18512.79 &10 & -18575.55 & 18 & -18655.49 \\
   3 & -18507.43 &11 & -18570.70 & 19 & -18645.04 \\
   4 & -18535.13 &12 & -18621.86 & 20 & -18635.40 \\
   5 & -18525.99 &13 & -18641.05 & 21 & -18626.54 \\
   6 & -18525.54 &14 & -18664.56 & 22 & -18620.68 \\
   7 & -18521.59 &15 & -18654.19 & 23 & -18613.30 \\
\hline \end{tabular}
\end{center}
\end{minipage}   
\hspace{5mm}
\begin{minipage}{0.35\textwidth}
\begin{center}
\makeatletter
\def\@captype{figure}
\includegraphics[width=50mm,angle=0,clip=]{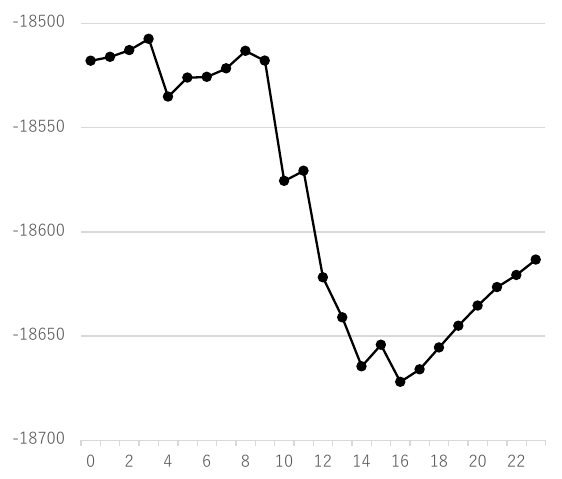}
\caption{Change of AIC of the two-periodic models for Toden data}
\label{Fig_Toden_AIC}
\end{center}
\end{minipage}
\end{table}

\begin{figure}[tbp]
\includegraphics[width=75mm,angle=0,clip=]{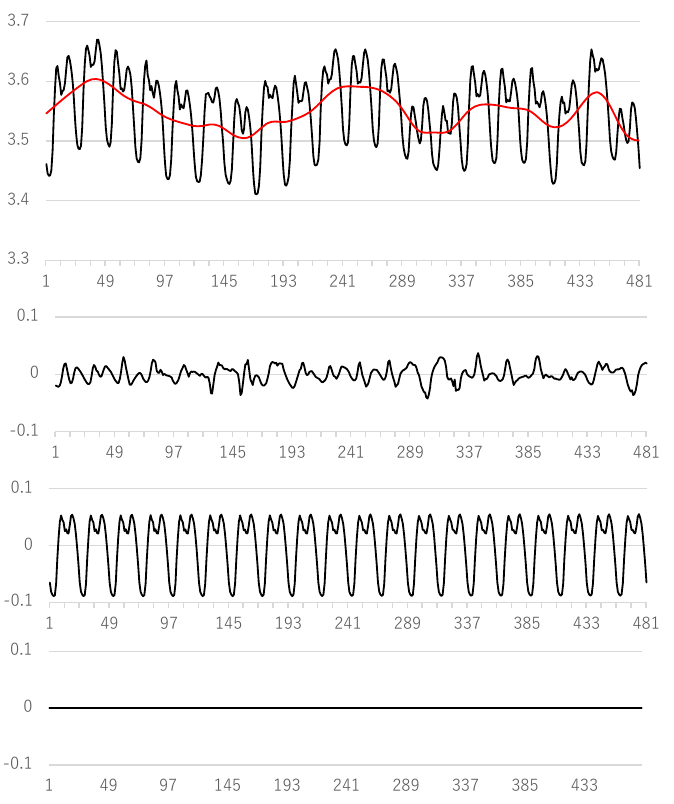}
\hspace{5mm}
\includegraphics[width=75mm,angle=0,clip=]{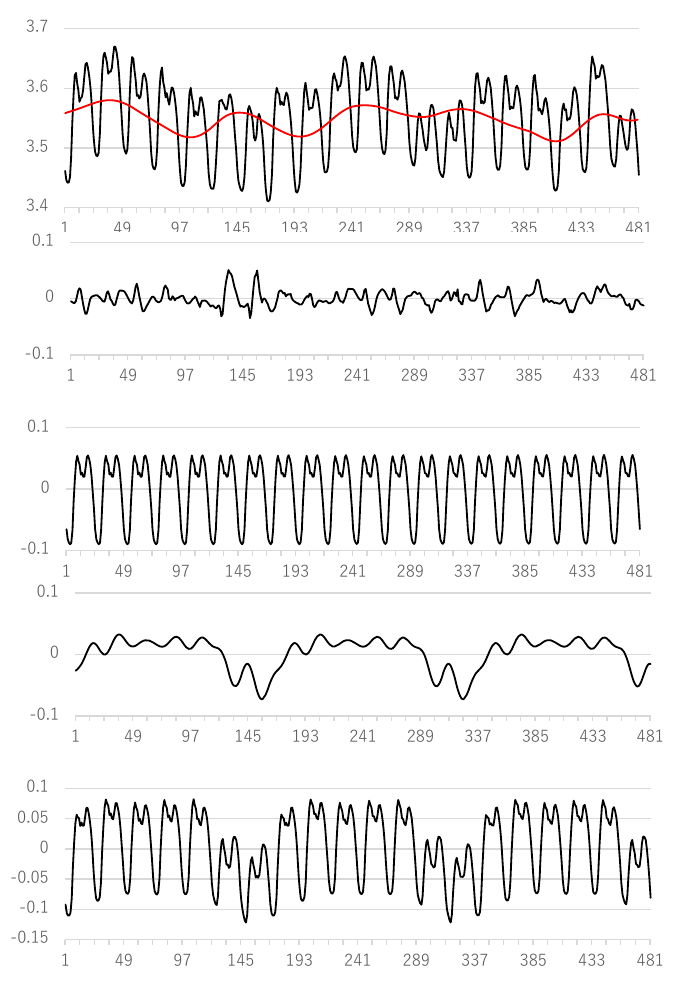}
\caption{Decomposition of Toden electricity data by standard Decomp model (left) and two-component seasonal model with $m_4=23$ and $m_5=16$}\label{Fig_Toden_Decomp}
\end{figure}

Left plots of Figure \ref{Fig_Toden_Decomp} show a decompsition by Decomp model with 
$m_1=2$, $m_2=1$ and $m_3=2$. Since the data is long, only the first 481 data are shown.
The seasonal component of Saturday and Sunday is different from the other days of the week, but the standard seasonal adjustment yields a constant seasonal component.

The right plots of the figure shows the decomposition using the two-period component model. 
From top to bottom, the observed data and trend, AR components, $Q_1$, $Q_2$ and $Q_1+Q_2$ are shown.
$Q_1$ is similar to the seasonal component of Decomp, but the weekly cycle component of $Q_2$ shows a significant decrease on weekends. As a result, the sum of two periodic components, $Q1+Q2$, shows a decrease in level and also a smaller variability on weekends.
The $Q_1+Q_2$ components shown at the bottom correspond well to the periodic variation of the original series

\begin{figure}[tbp]
\begin{center}
\includegraphics[width=120mm,angle=0,clip=]{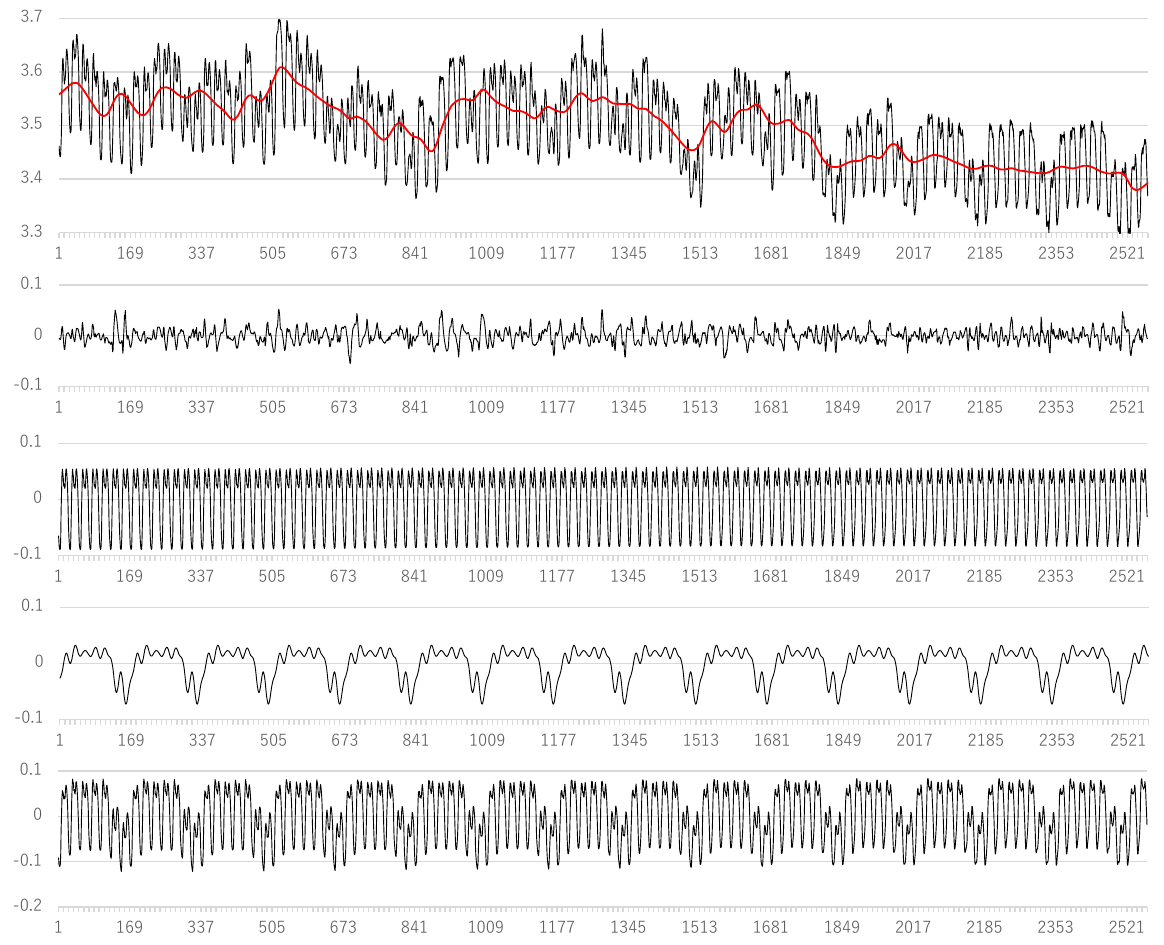}
\caption{Decomposition of entire Toden electricity data by two-component seasonal model with $m_4=23$ and $m_5=16$}\label{Fig_Toden_2-cycle_M5=16}
\end{center}
\end{figure}

Figure \ref{Fig_Toden_2-cycle_M5=16} shows the results of decomposition of the entire time series by the same model as shown in right plots of Figure \ref{Fig_Toden_Decomp}. The first 481 points are identical to those in Figure \ref{Fig_Toden_Decomp}.

\begin{figure}[tbp]
\includegraphics[width=74mm,angle=0,clip=]{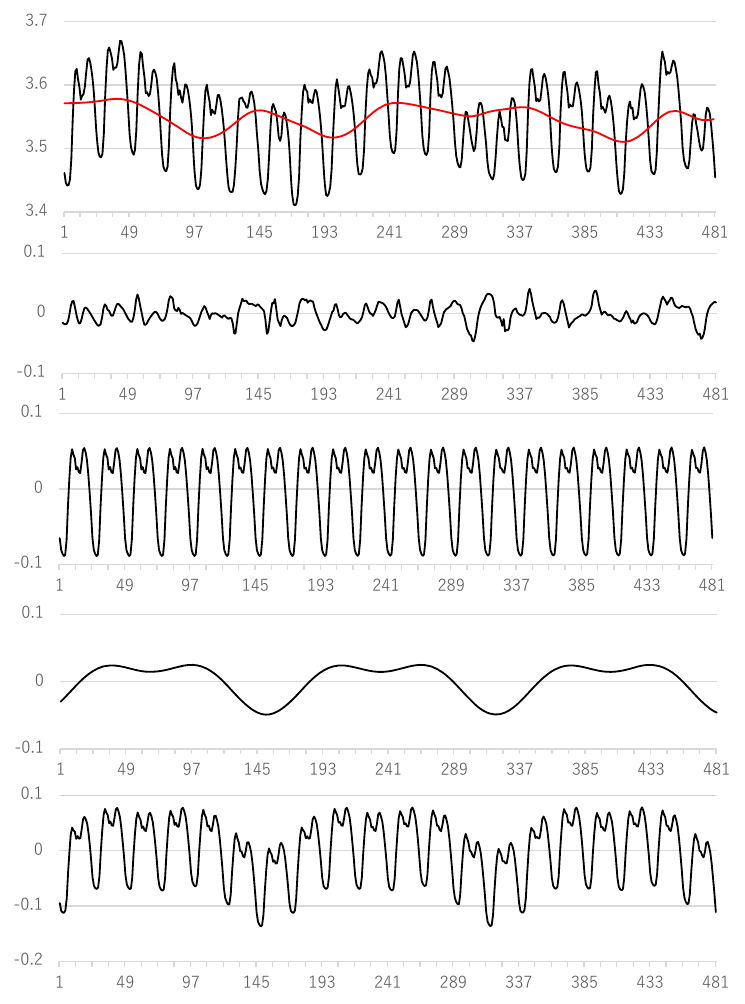}
\hspace{5mm}
\includegraphics[width=76mm,angle=0,clip=]{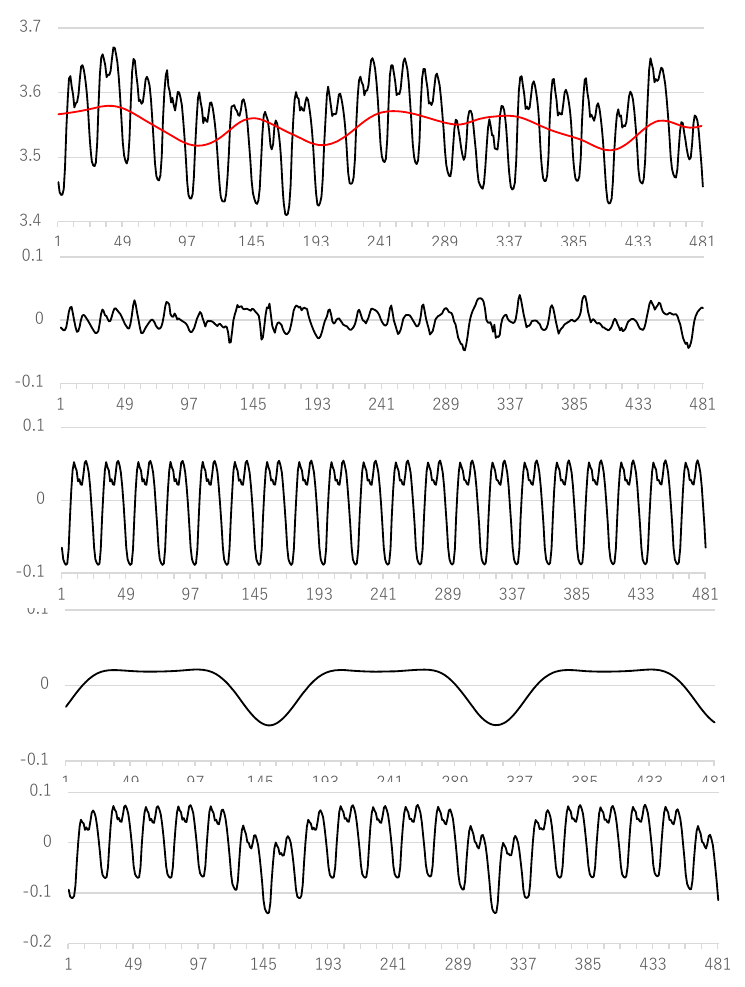}
\caption{}\label{Fig_Toden_2-cycle_M5=4,6}
\end{figure}

Figure \ref{Fig_Toden_2-cycle_M5=4,6} shows the change in the $Q_2$ component when a smaller $m_5$ is used. The left figure is for $m_5$ = 4 and the right is for $m_5$ = 6. In these cases, the detailed pattern of variation are different, but the lower electricity consumption on weekends is captured, resulting in no significant difference in the trend estimates.


\newpage
\noindent
\section{Seasonal Adjustment of Time Series with Long and Short Seasonality}

In the previous example, trigonometric seasonal models were used for both of the two seasonal components, but a Decomp model seasonal component model could be used for seasonal components with shorter periods.

Here we consider a time series with two types of seasonality, one with a very long period
such as the annual cycle and the other with a short period such as the daily cycle.
To treat such a time series with two types of seasonalities, we consider the following model
\begin{eqnarray}
y_n = t_n + S_n + p_n + Q_n + w_n 
\label{Eq_LS_obs_model}
\end{eqnarray}
where $y_n$ is time series, and $t_n$, $S_n$, $Q_n$, $p_n$ and $w_n$ are
the trend component, the seasonal component with short period, the seasonal component with long period,
stationary AR component and the noise component, respectively.

These components are assumed to follow 
\begin{eqnarray}
&& t_n = t_{n-1} + v_n^{(t)},\quad {\rm or}\quad t_n = 2t_{n-1} - t_{n-2} + v_n^{(t)} \nonumber \\
&& S_n = -(S_{n-1} + \cdots + S_{n-p+1}) + v_n^{(s)}\nonumber \\
&& Q_n = \sum_{j=1}^{k_c} c_{j,n} \cos (\omega jn) + \sum_{j=1}^{k_s} d_{j,n} \sin (\omega jn) \\
\label{Eq_LS_componennt_model}
&&p_n = \sum_{j=1}^{m_3} a_j p_{n-j}  + v_n^{(p)}, \nonumber 
\end{eqnarray}
where $m_4=k_c+k_s$ and if $m_4$ is a even number $k_c=k_s= \frac{m_4}{2}$ and if $m_4$ is a odd
number, $k_c=\frac{m_4}{2}+1$ and $k_s=\frac{m_4}{2}$. For the coefficients $c_{j,n}$ and $d_{j,n}$, we considere the following two types of models. 
The first is the constant coefficient model for which $c_{j,n} = c_j$ and 
$d_{j,n}=d_j$ hold, and the the second is a random walk models that follow
\begin{eqnarray}
c_{j,n} &=& c_{j,n-1} + v_{j,n}^{(c)} \nonumber \\
d_{j,n} &=& d_{j,n-1} + v_{j,n}^{(d)}.
\end{eqnarray}
The noise components $v_{j,n}^{(c)}$, $v_{j,n}^{(d)}$ and $w_n$ are assumed to follow Gaussian white noise with mean 0 and the variance $\tau_q^2$, $\tau_q^2$ and $\sigma^2$, respectively. 
Note that for simplicity, it is assumed that all of $v_{j,n}^{(c)}$ and $v_{j,n}^{(d)}$, $j=1,\ldots ,k$ have a common variance $\tau_q^2$.

Assuming that $m_1=1$ and $m_3=2$, the basic model for time series with long and short periods (\ref{Eq_LS_obs_model})
and the component models (\ref{Eq_LS_componennt_model}) can be expressed in the 
state-space model form
\begin{eqnarray}
\arraycolsep=0mm
 x_n &=& \left[ \begin{array}{c}
          S_n \\ S_{n-1} \\ \vdots \\ S_{n-11} \\ p_n \\ p_{n-1}\\
          d_0 \\ c_1 \\ d_1 \\ \vdots \\ d_k \end{array}\right], \quad
\arraycolsep=1mm
F = \left[ \begin{array}{cccc|cc|ccccc}
       -1 &-1 &\cdots&-1& & & & & & & \\
       1  &   &      &  & & & & & & & \\
          &\ddots&   &  & & & & & & & \\
          &   &  1&  &  & & & & & &  \\ \hline
          &   &   &  &a_1&a_2&& & & &  \\
          &   &   &  &  1& 0 & & & & &  \\ \hline
          &   &   &  &   &   & 1& & & &  \\
          &   &   &  &   &   &  &1& & &  \\
          &   &   &  &   &   &  & &1& &  \\
          &   &   &  &   &   &  & & &\ddots&  \\
          &   &   &  &   &   &  & & &      & 1 \end{array}\right],\quad
G = \left[ \begin{array}{c|c|ccccc}
        1 &  &  &  &  & & \\
        0 &  &  &  &  & &\\
    \vdots&  &  &  &  & &\\
        0 &  &  &  &  & &\\ \hline
          & 1&  &  &  & &\\
          & 0&  &  &  & &\\ \hline
          &  & 1&  &  & &\\
          &  &  & 1&  & &\\
          &  &  &  & 1& &\\
          &  &  &  &  & \ddots& \\
          &  &  &  &  &  & 1\\
 \end{array}\right], \nonumber\\
H_n&=&\left[ \begin{array}{cccc|cc|cccccc} 
        1 & \cdots & 0 & 1 &0 &  1 & 1 &\cos(\omega n) & \sin (\omega n) & \cdots & \cos(\omega kn) & \sin (\omega kn)\end{array}\right] \\
Q &=& \left[ \begin{array}{cc|ccc}
           \tau^2_{s}&          &          & & \\
                     &\tau^2_{p}&          & &  \\ \hline
                     &          &\tau^2_{q}& &  \\
                     &          &          & \ddots & \\ 
                     &          &          &   &\tau^2_{q} 
               \end{array}\right],
\quad R = \sigma^2.
\end{eqnarray}
For the constant coefficients model, $\tau^2_{q}$ is set to be zero.
Note that in this case, the trend component is included in

The log-likelihood of the model can be computed using this state-space model
and the unknown parameters of the model such as the variances of the 
noises, $\tau^2_s$, $\tau^2_p$, $\tau^2_{q}$, $\sigma^2$ and the
autoregressive coefficents, $a_1,\ldots ,a_{m_3}$ are estimated
by maximizing the log-likelihood function.
In computing the AIC, the number of parameters is the sum of the
one in the trigonometric regression model and the one in the 
DECOMP model, i.e., $k+1 + id(m_1) + id(m_2) + id(m_3) + 1 + m_3$.
\begin{eqnarray}
{\rm AIC} = -2 \ell + 2\{2k+1 + id(m_1) + id(m_2) + id(m_3) + 1 + m_3\}.
\end{eqnarray}

\subsection{Example:  Temperature data}

We considere here the temperature data observed from 15:00 November 19, 2011 to 13:00 January 11, 2014, at Narusawa, Yamanashi prefecture, Japan.
The original record was observed every 30 minuits, but here it is resampled at each 2 hours.
Therefore, the data length is $N=18900$. 
The data has both a daily seasonality ($p=12$) and an annual seasonality ($q=4380$).

\subsubsection{Two-step modeling}
Besides the two type of long-period seasonal component models, we also consider a
two-step method, namely, we first delete the long-period seasonal component by
fitting and removing trigonometric regression component
\begin{eqnarray}
z_n = y_n - Q_n^{(k)}.
\end{eqnarray}
We then fit an ordinary seasonal adjustment model
\begin{eqnarray}
z_n = t_n + S_n + p_n + w_n,  \quad w_n \sim N(0,\sigma^2).
\end{eqnarray}
The AIC of the model is obtained by
\begin{eqnarray}
{\rm AIC} = -2 \ell + 2\{2k+1 + id(m_1) + id(m_2) + id(m_3) + 1 + m_3 \}.
\end{eqnarray}
where $2k+1$ is the number of the trigonometric regression coefficents.

In the actual data analysis, the trigonometric regression models up to the order $k=182$ were considered.
The highest order model has 365 unknown coefficients, one constant term and
182 sine and cosine components.
The minimum of AIC, 106606.84, was attained at $k=102$.
However, since this model contains 205 regression coefficients
and the additional penalty term in the modelified AIC becomes
very large, we also consider an arbitralily selected lower order $k=7$, 
for the first step.
We obtained an annual seasonality removed time series by 
deleting the trigonometric regression with order $k=7$ and $k=102$.

\begin{figure}[tbp]
\begin{center}
\includegraphics[width=150mm,angle=0,clip=]{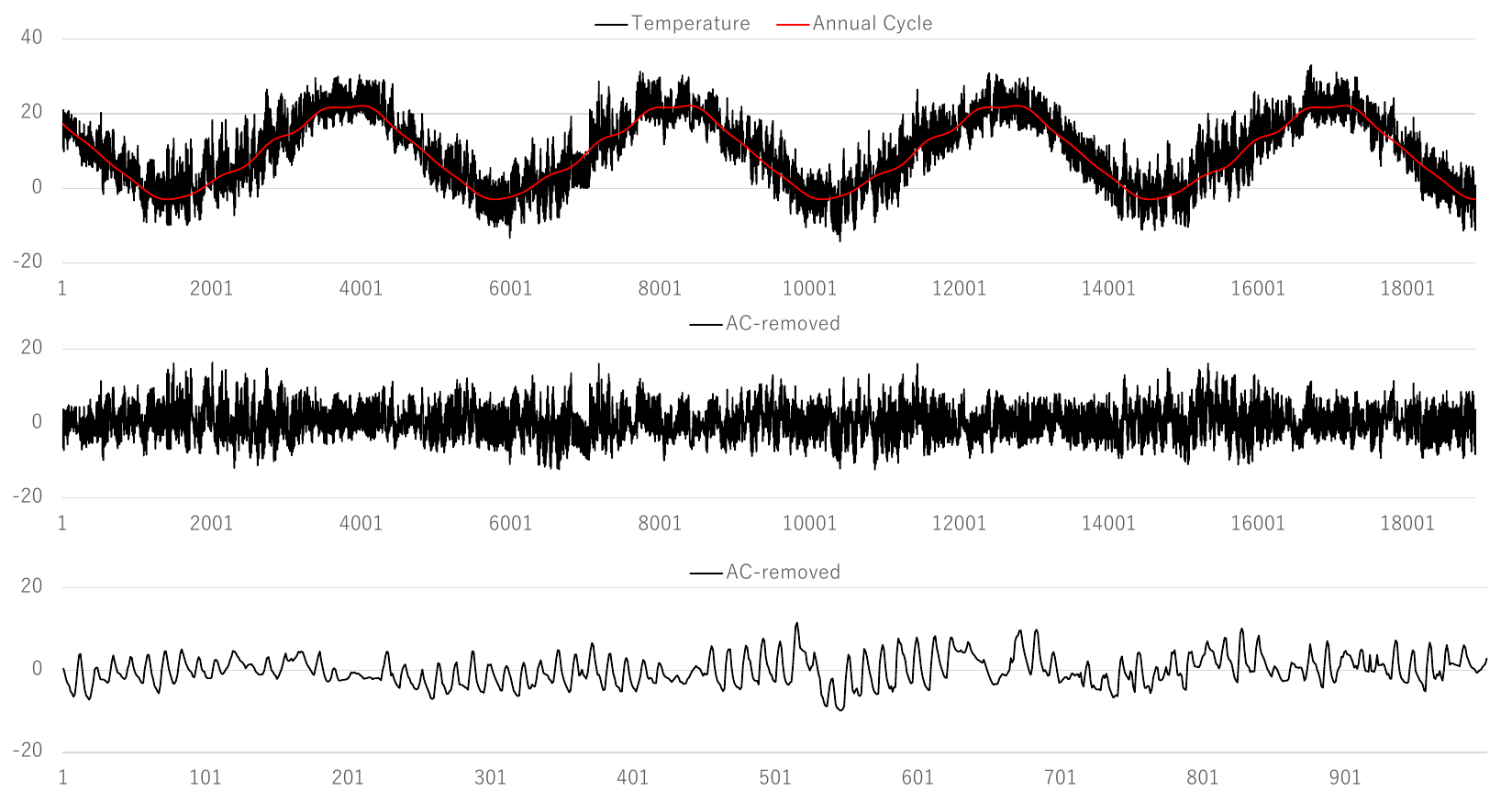}
\end{center}
\caption{Top plot: Temperature and estimated annual cycle by a trigonometric regression model with $k=7$. Middle plot: Annual cycle removed temerature series. Bottom plot: The first 1000 series of Annual cycle removed data.}\label{Fig_annual_cycle}
\end{figure}

Figure \ref{Fig_annual_cycle} shows the decomposition of 
the original temperature data.
The top plot shows the original temperature data and the estimated
annual cycle by the trigonometric regression curve with $k=7$,
and the second plot shows the series obtained by removing the 
estimated annual cycle from the original data.
Hereafter, we referred to this series as the annual cycle removed
temperature data.
The bottom plot shows the first 1000 data of the annual cycle removed temperature data.
Since this data is 2-hour interval data, the 1-day cycle ($p$ = 12) becomes pronounced.

\begin{table}[bp]
\begin{center}
\caption{The log-likelihood and AIC of the models with various orders for Temperature data}\label{Tab_AIC_temperature}

\vspace{3mm}
\begin{tabular}{c|ccc|ccc}
          &  &$k=7$  &  &  & $k=102$ &    \\
    $m_3$ & log-likelihood & {\rm AIC}& {\rm AIC}${}^{\prime}$& log-likelihood & {\rm AIC} & {\rm AIC}${}^{\prime}$\\
\hline
     0 & -47864.60 & 95733.20 & 95763.20 & -47856.95 & 95717.89 & 96127.89 \\
     1 & -31699.38 & 63406.76 & 63436.76 & -31313.11 & 62634.22 & 63044.22 \\
     2 & -31462.27 & 62934.54 & 62964.54 & -31019.41 & 62048.83 & 62458.83 \\
     3 & -31455.53 & 62923.07 & 62953.06 & -30999.73 & 62011.45 & 62421.45 \\
     4 & -31431.93 & 62877.87 & 62907.87 & -30975.29 & 61964.58 & 62374.58 \\
     5 & -31398.58 & 62813.15 & 62843.15 & -30934.28 & 61884.56 & 62294.56 \\
     6 & -30800.39 & 61618.78 & 61648.78 & -30870.76 & 61759.53 & 62169.53 \\ 
     7 & -30705.09 & 61430.19 & 61460.19 & -30840.37 & 61700.74 & 62110.74 \\ 
     8 & -30694.47 & 61410.93 & 61440.93 & -30753.10 & 61528.19 & 61938.19 \\ 
     9 & -30667.49 & 61358.99 & 61388.99 & -30702.00 & 61428.00 & 61837.00  \\ 
    10 & -30509.91 & 61045.83 & 61075.83 & -30694.43 & 61414.86 & 61824.86  \\ 
    11 & -30495.10 & 61018.19 & 61048.19 & -30674.30 & 61376.60 & 61786.60  \\ 
    12 & -30494.15 & 61018.29 & 61048.29 & -30619.41 & 61268.81 & 61678.81  \\ 
    13 & -30489.42 & 61010.84 & 61040.84 & -30495.52 & 61023.05 & 61433.05  \\ 
    14 & -30486.71 & 61007.41 & \textbf{61037.41} & -30492.22 & 61018.44 & 61428.44  \\ 
    15 & -30486.38 & 61008.76 & 61038.76 & -30489.29 & 61014.58 & \textbf{61424.58}  \\ 
  \hline
\end{tabular}
\end{center}
\end{table}

Table \ref{Tab_AIC_temperature} shows the log-likelihoods and AIC's of the 
seasonal adjustment models with stationary AR component.
AIC${}^{\prime}$ shows the AIC's of the model with respect to the original
data before removing the annual cycle.
The table shows the two cases, i.e., $k=7$ and $k=102$.
Since the models for $k=102$ contains 205(=$2k+1$) additional parameters,
AIC is significantly  larger than the ones for $k=7$.
Within the models with $k=7$, the model with AR order $m_3=14$ attains the
minimum of the AIC${}^{\prime}$.

\begin{figure}[tbp]
\begin{center}
\includegraphics[width=120mm,angle=0,clip=]{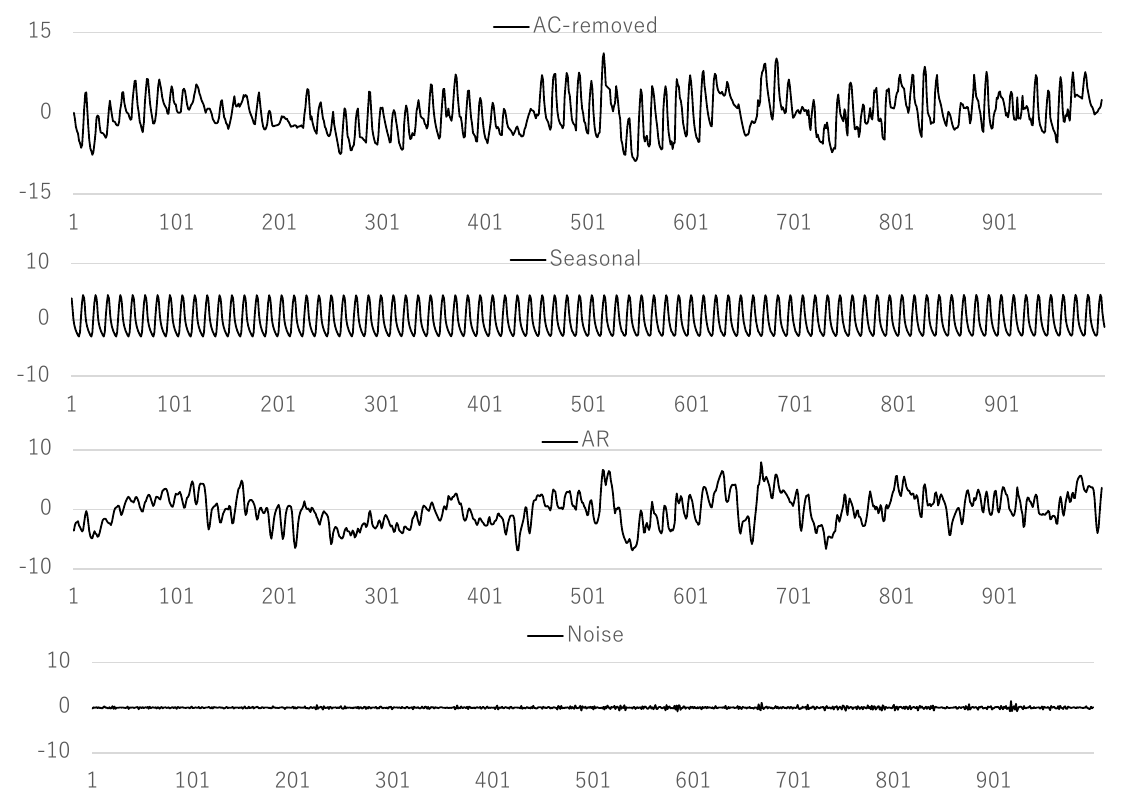}
\end{center}
\caption{Decomposition of the first part of the annual cycle-removed temperature data by the conventional seasonal adjustment model with, $m_1=0$, $m_2=1$, $m_3=15$. $k=7$ was use in removing the annual cycle.
}\label{Fig_DECOMP_AC-removed_temperature_part}
\end{figure}

\begin{figure}[tbp]
\begin{center}
\includegraphics[width=150mm,angle=0,clip=]{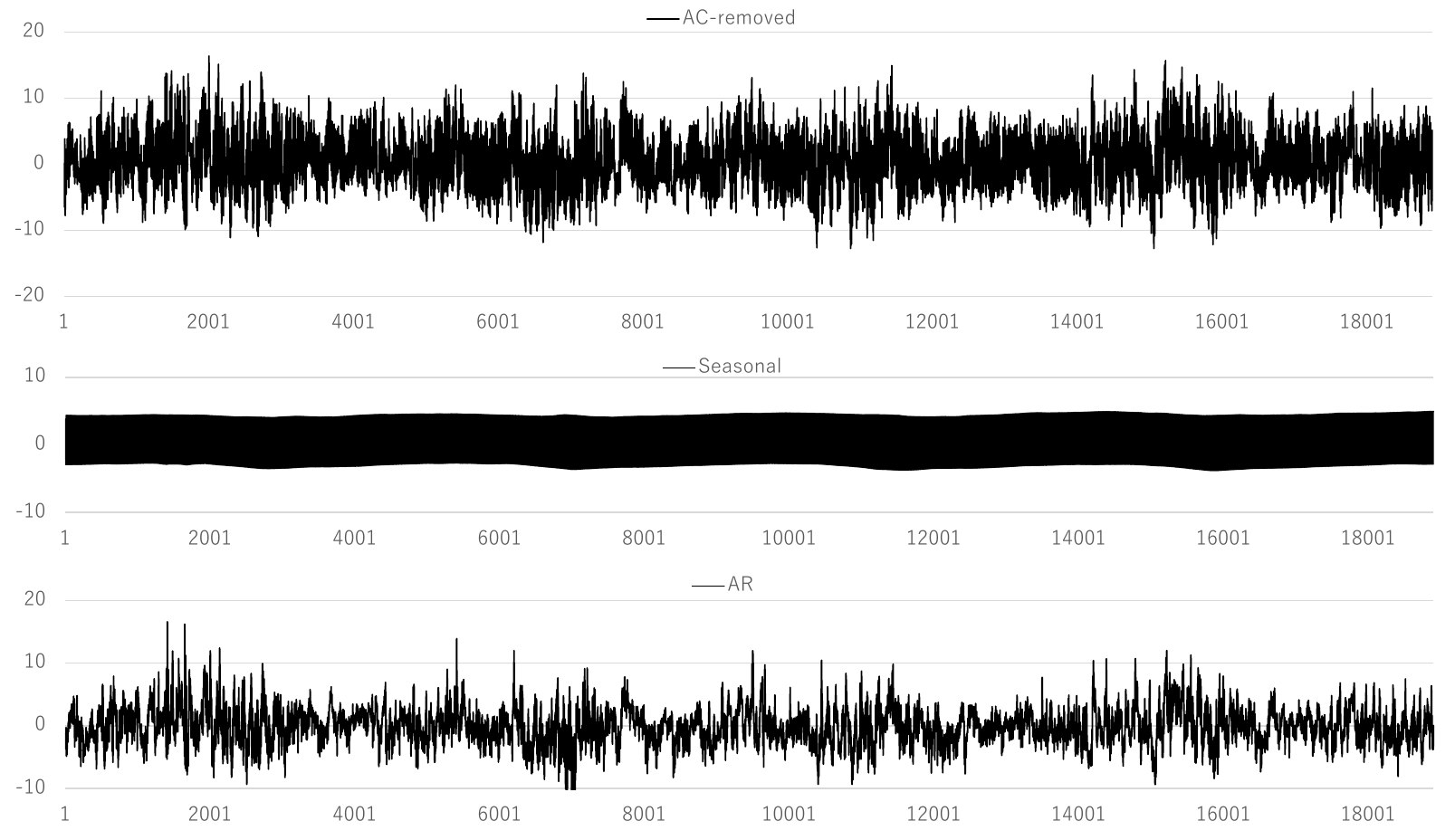}
\end{center}
\caption{Decomposition of annual cycle-removed temperature data by the conventional seasonal adjustment model with, $m_1=0$, $m_2=1$, $m_3=15$. $k=7$ was use in removing the annual cycle.
}\label{Fig_DECOMP_AC-removed_temperature}
\end{figure}

Figure \ref{Fig_DECOMP_AC-removed_temperature_part} shows the results of 
seasonal adjustment of the first 1000 annual cycle removed temerature data by the
seasonal adjustment model with $m_1$=0, $m_2=1$ and $m_3=15$.
From top to bottom, the four plots show, annual cycle removed temerature data,
estimated daily cycle, AR process and the noise component.
Figure \ref{Fig_DECOMP_AC-removed_temperature} shows the decomposition of the 
entire data obtained by this two-step modeling.

\subsubsection{One-factor annual cycle model}
Hereafter, we consider direct modeling of temperature data without
removing annual cycle.
We first consider one-factor model with the annual seasonal component
is expressed as
\begin{eqnarray}
y_n &=& t_n + S_n + p_n + Q_n^{(k)} + w_n  \nonumber \\
t_n &=& t_{n-1} + v_n^{(T)} \quad {\rm or}\quad t_n = 2t_{n-1} - t_{n-2}  + v_n^{(T)} \nonumber \\
S_n &=& -(S_{n-1}+\cdots + S_{n-p+1}) + v_n^{(s)}  \\
p_n &=& \sum_{j=1}^{m_3} a_j p_{n-j} + v_n^{(p)} \nonumber \\
Q_n^{(k)} &=& \beta_n \left\{  \sum_{j=1}^k c_{j} \sin (\omega jn) + \sum_{j=1}^k d_{j} \cos (\omega jn)  \right\} \nonumber
\end{eqnarray}
where $\omega=\frac{2\pi}{4380}$ and the common factor $\beta_n$ follows a random walk model
\begin{eqnarray}
\beta_n = \beta_{n-1} + v_n^{(\beta)},\quad v_n^{(\beta)} \sim N(0,\tau^2_{\beta}).
\end{eqnarray}
The order $k$ is selected as 15 and two types of trend order $m_1=1$ and 2 are
considered. AR order up to $m_3=18$ was considered.
From Table \ref{Tab_AIC_one-factor_models}, it can be seen that the model with
$m_1=1$, $m_3=15$ attains the minimum of AIC.

\begin{table}[tbp]
\begin{center}
\caption{The log-likelihood and AIC of the one-factor annual cycle models with various orders for Temperature data}\label{Tab_AIC_one-factor_models}
\begin{tabular}{c|cc|cc}
         & \multicolumn{2}{c|}{$m_1=1$} & \multicolumn{2}{c}{$m_1=2$}   \\
$m_3$ & log-likelihood & {\rm AIC}& log-likelihood & {\rm AIC} \\
\hline
    0 & -31763.61 & 63555.22 &-34696.66 & 69423.32 \\
    1 & -31394.69 & 62819.39 &-31404.40 & 62840.81 \\
    2 & -31049.47 & 62130.93 &-31120.21 & 62274.41 \\
    3 & -31044.81 & 62123.61 &-31103.82 & 62243.64 \\
    4 & -30916.39 & 61868.78 &-31076.27 & 62190.54 \\
    5 & -30911.69 & 61861.37 &-31033.22 & 62106.44 \\
    6 & -30908.80 & 61857.60 &-30970.45 & 61982.90 \\
    7 & -30834.66 & 61711.32 &-30942.29 & 61928.57 \\
    8 & -30728.28 & 61500.55 &-30863.45 & 61772.91 \\
    9 & -30700.85 & 61447.71 &-30821.00 & 61690.00 \\
   10 & -30611.30 & 61270.61 &-30817.00 & 61684.01 \\
   11 & -30592.06 & 61234.13 &-30804.45 & 61660.91 \\
   12 & -30589.01 & 61230.03 &-30737.74 & 61529.49 \\
   13 & -30584.73 & 61223.46 &-30599.77 & 61255.54 \\
   14 & -30583.70 & 61223.41 &-30594.09 & 61246.17 \\
   15 & -30580.54 & \textbf{61219.07} &-30592.53 & \textbf{61245.07} \\
   16 & -30580.42 & 61220.84 &-30592.37 & 61246.75 \\
   17 & -30580.40 & 61222.81 &-30592.32 & 61248.65 \\
   18 & -30580.10 & 61224.20 &-30591.82 & 61249.64 \\
   \hline
\end{tabular}
\end{center}
\end{table}

\begin{figure}[tbp]
\begin{center}
\includegraphics[width=120mm,angle=0,clip=]{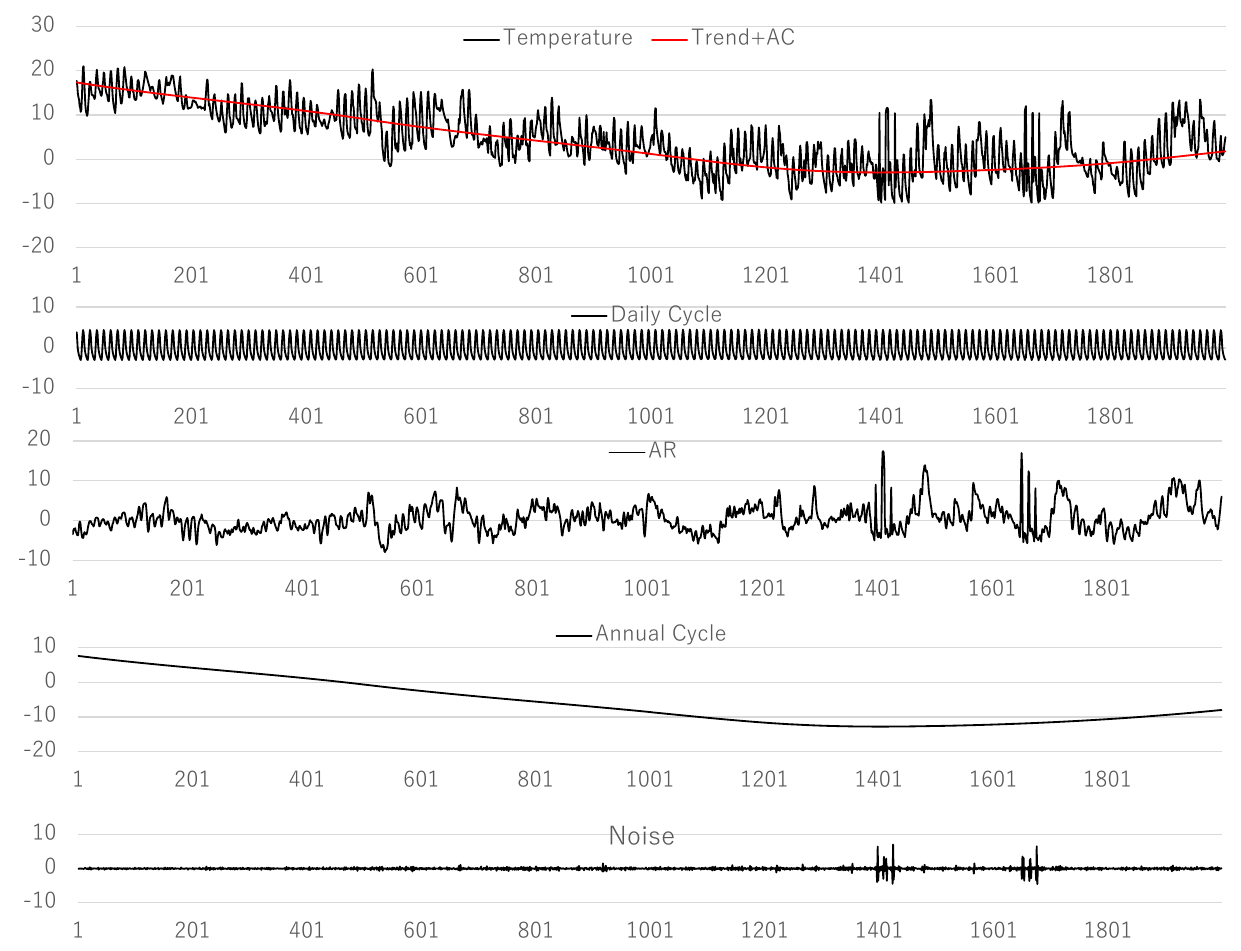}
\end{center}
\caption{Decomposition of the first part of the temperature data by the seasonal adjustment model with one-factor time-varying coefficient annual cycle models, $m_1=1$, $m_2=1$, $m_3=15$ and $m_4=15$.
}\label{Fig_one-factor_part}
\end{figure}

\begin{figure}[tbp]
\begin{center}
\includegraphics[width=150mm,angle=0,clip=]{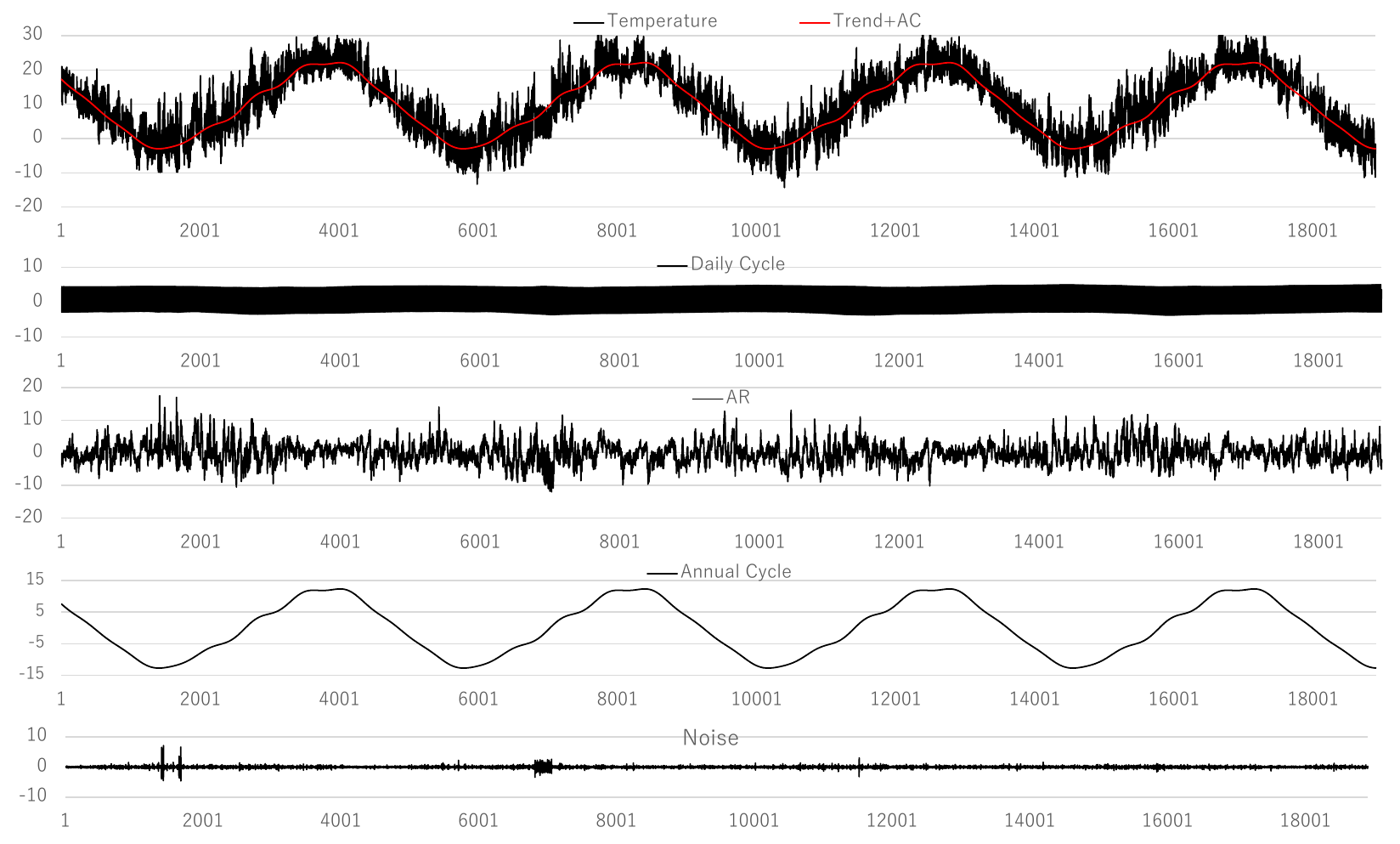}
\end{center}
\caption{Decomposition of temperature data by the seasonal adjustment model with one-factor time-varying coefficient annual cycle models, $m_1=1$, $m_2=1$, $m_3=15$ and $m_4=14$.
}\label{Fig_one-factor}
\end{figure}

Figure \ref{Fig_one-factor_part} shows the decompsition of the first part of the
temperature data ($n$=1000) by the one-factor annual cycle model.
Although we consider a time-varying one factor annual cycle, in this AIC-best model, 
the estimated variance of the factor $\beta_n$ is virtually zero and the estiamted
factor $\beta_n$ is actually time-invariant.
Figure \ref{Fig_one-factor} shows the decompsition of the whole temperature data 
($n$=18900) by the same one-factor annual cycle model.

\subsubsection{Time-varying coefficient annual cycle model}

We consider a time-varying coefficients annual cycle model,
where the time series is decomposed as
\begin{eqnarray}
y_n &=& S_n + Q_n^{(k)} + p_n + w_n  \nonumber \\
S_n &=& S_{n-p} + v_n^{(s)} \nonumber \\
p_n &=& \sum_{j=1}^{m_3} a_j p_{n-j} + v_n^{(p)}  \\
Q_n^{(k)} &=& \beta_n \left\{  \sum_{j=0}^k c_{j} \sin (\omega jn) + \sum_{j=1}^k d_{j} \cos (\omega jn)  \right\} \nonumber
\end{eqnarray}
Here, since this model does not contain a trend component, it is
possible to use random walk type daily seasonal model instead 
of the one used in the previous models.
Here, this model does not include a trend component, but the first term ($\sin(0)$) of $Q_n^{(k)}$ component plays the same role as the random walk type trend component of $m_1$=1.

Table \ref{Tab_AIC_temperature_TVC-model} shows the change of the log-likelihood
and AIC when various models with the number of sinusoidal components
$m_4=3$, 7, 11 and the order of autoregressive components, $m_3=1,\ldots ,18$.
From the table, it can be seen that model with $m_3=14$ and $m_4=11$
attains the minimum of AIC.

\begin{figure}[tbp]
\begin{center}
\includegraphics[width=120mm,angle=0,clip=]{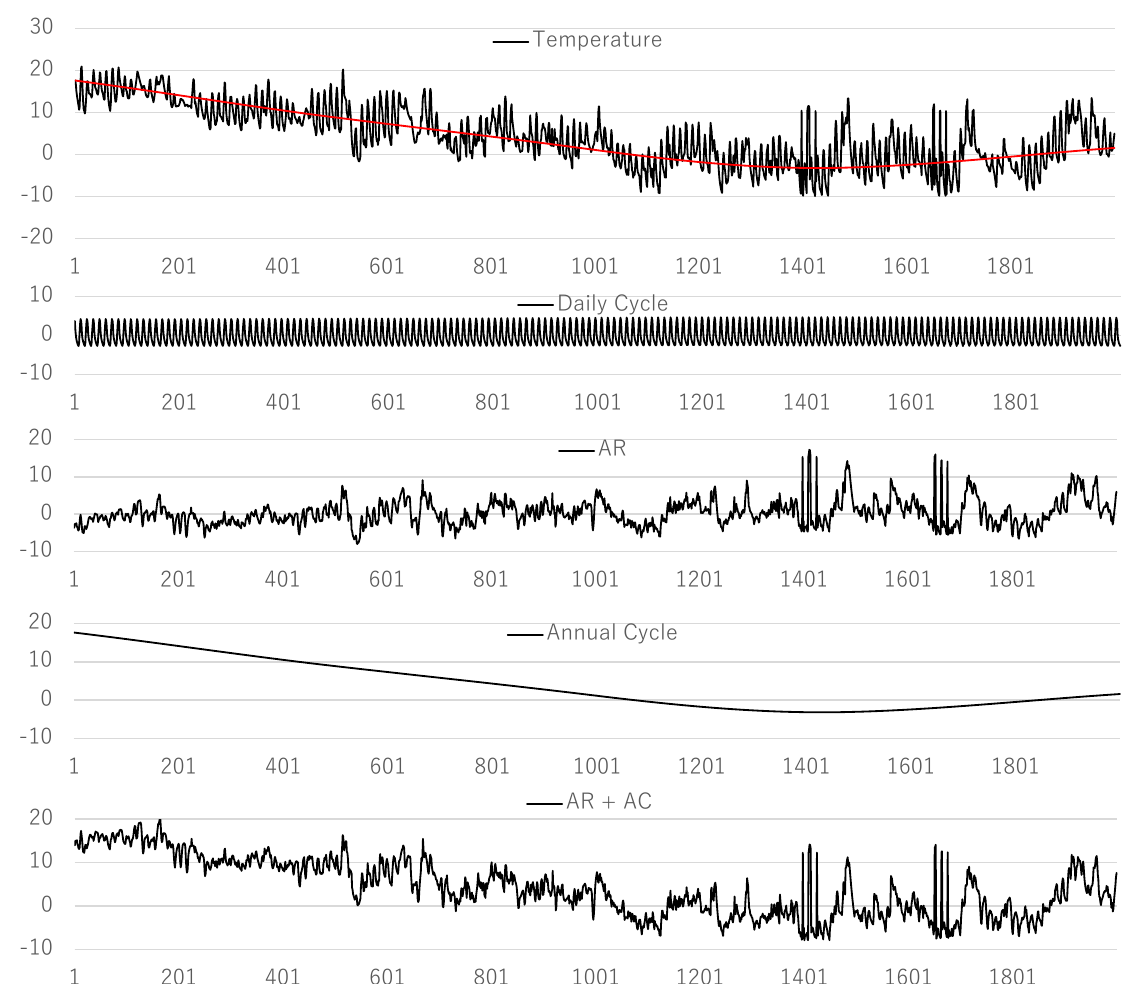}
\end{center}
\caption{Decomposition of the first part of the temperature data by the seasonal adjustment model with time-varying coefficient annual cycle models, $m_1=0$, $m_2=1$, $m_3=14$ and $m_4=11$.
}\label{Fig_temperature_011411_part}
\end{figure}

\begin{figure}[tbp]
\begin{center}
\includegraphics[width=150mm,angle=0,clip=]{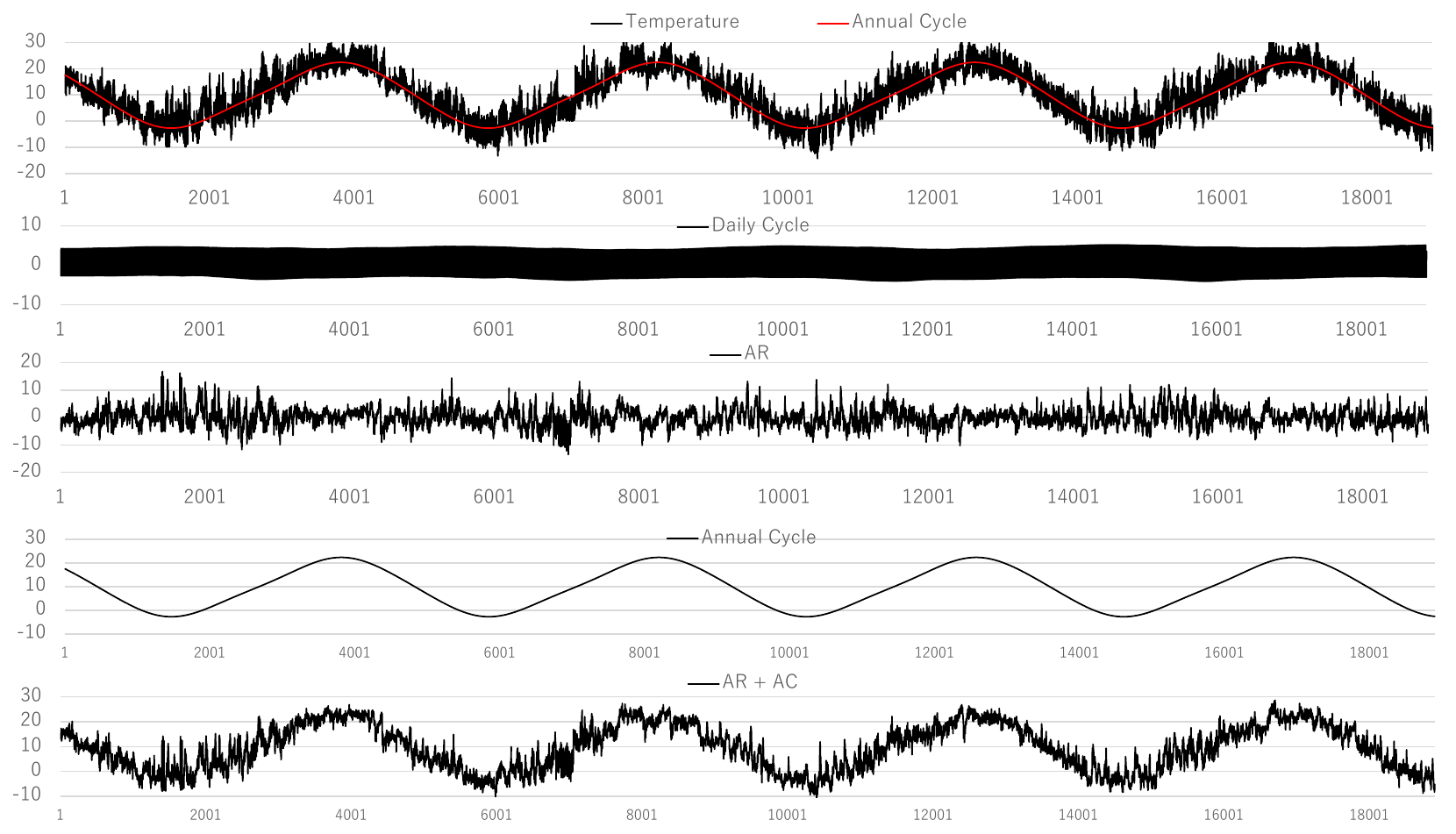}
\end{center}
\caption{Decomposition of temperature data by the seasonal adjustment model with time-varying coefficient annual cycle models, $m_1=0$, $m_2=1$, $m_3=14$ and $m_4=11$.
}\label{Fig_temperature_011411}
\end{figure}

Figure \ref{Fig_temperature_011411_part} shows the decomposition of the first part 
of the temperature data.
It can be seen that the visually similar results to the previous one-factor model
was obtained with this model.
Figure \ref{Fig_temperature_011411} shows the decomposition of the entire temperature data.
Compared to Figure \ref{Fig_one-factor}, the annual cycle is simpler. 
This may be due to the higher order of $k=1$5 ($m_4=30$) in the case of Figure \ref{Fig_one-factor}.

\begin{table}[h]
\begin{center}
\caption{The log-likelihood and AIC of the time-varying coefficients annual cycle models with various orders for Temperature data. No trend comonent, $m_1=0$.}\label{Tab_AIC_temperature_TVC-model}
\begin{tabular}{c|cc|cc|cc}
         & \multicolumn{2}{c}{$m_4=3$} & \multicolumn{2}{|c|}{$m_4=7$} & \multicolumn{2}{c}{$m_4=11$} \\
   $m_3$ & log-likelihood & {\rm AIC}& log-likelihood & {\rm AIC}& log-likelihood & {\rm AIC}\\
\hline
    1 & -31245.52 & 62501.04 &-31229.64 & 62469.28 &-31227.14 & 62464.27 \\
    2 & -31019.69 & 62051.39 &-30997.77 & 62007.53 &-30994.01 & 62000.03 \\
    3 & -31006.55 & 62027.10 &-30981.84 & 61977.68 &-30979.06 & 61972.12 \\
    4 & -30978.11 & 61972.22 &-30956.81 & 61929.63 &-30953.46 & 61922.93 \\
    5 & -30933.30 & 61884.60 &-30914.77 & 61847.54 &-30911.80 & 61841.61 \\
    6 & -30867.56 & 61755.12 &-30851.84 & 61723.05 &-30849.32 & 61718.63 \\
    7 & -30839.62 & 61701.23 &-30825.52 & 61673.05 &-30823.26 & 61668.52 \\
    8 & -30764.76 & 61553.52 &-30752.96 & 61529.93 &-30751.06 & 61526.12 \\
    9 & -30728.45 & 61482.90 &-30718.06 & 61462.11 &-30716.37 & 61458.74 \\
   10 & -30726.71 & 61481.42 &-30716.59 & 61440.78 &-30714.94 & 61457.89 \\
   11 & -30714.81 & 61459.63 &-30705.39 & 61328.25 &-30703.85 & 61437.70 \\
   12 & -30659.10 & 61350.20 &-30648.13 & 61092.05 &-30646.35 & 61324.70 \\
   13 & -30542.70 & 61119.40 &-30529.03 & 61086.98 &-30526.84 & 61087.68 \\
   14 & -30539.69 & 61115.38 &-30525.49 & \textbf{61086.92} &-30523.22 & \textbf{61082.43} \\
   15 & -30538.33 & \textbf{61114.66} &-30524.46 & 61087.83 &-30522.23 & 61082.46 \\
   16 & -30537.98 & 61115.96 &-30523.91 & 61087.82 &-30523.78 & 61087.56 \\
   17 & -30537.85 & 61117.69 &-30523.86 & 61089.72 &-30521.65 & 61085.30 \\
   18 & -30537.76 & 61119.52 &-30523.67 & 61091.33 &-30521.39 & 61086.78 \\
   \hline
\end{tabular}
\end{center}
\end{table}

\newpage
\section{Concluding Remarks}

In this paper, the seasonal adjustment method using trigonometric functions is examined on the basis of actual cases, and the following findings are made.
\begin{itemize}
\item When there is only one seasonal component and the period is short, there is no significant difference from the Decomp model. However, when the seasonal variation pattern is simple and can be expressed with a small number of components, there is some advantage.
\item When there are multiple seasonal components, there is a significant advantage in using the trigonometric model for seasonal components with long periods, such as the annual cycle. This is because even in the case of a very long period, it can often be expressed with a relatively small number of trigonometric functions. In this case, unlike the Decomp type model, a very high order state vector is not required.
\item In the two seasonal components model, for seasonal component with short period, there seems to be no significant difference between the Decomp model and the trigonometric model.
\item When a Decomp type model is used for both of the two seasonal components, it is necessary and required for the model to guarantee the uniqueness of the decomposition (Haba and Nagao (2018)).
As shown in Section 3, when using the trigonometric model for the long-period seasonal component, it is easy to handle the case where there is a $f_2 = k f_1$ relationship between the two periods, since it is only necessary to exclude frequency components that are multiples of $k$.
\end{itemize}

\vspace{15mm}
%


\begin{thebibliography}{2}

\item Akaike, H. (1980b), ^^ ^^ Seasonal adjustment by a Bayesian modeling", {\it J. Time Series Anal.}, {\bf 1}, 1--13.

\item Akaike, H. and Ishiguro, M. (1983), \lq\lq Comparative study of X-11 and Bayesian procedure of seasonal adjustment," {\it Applied Time Series Analysis of Economic Data}, U.S. Census Bureau.

\item Box, G.E.P., Hillmer, S.C. and Tiao, G.C. (1978), ^^ ^^ Analysis and modeling of seasonal time series", in {\it Seasonal Analysis of Time Seres}, ed.Zellner, A., US Bureau of the Census, {\it Economic Research Report ER-1}, 309--334.

\item Cleveland W. S. and Tiao G. C. (1976), ^^ ^^ Decomposition of seasonal time series: a model for the Census X-11 program", \textit{J. Am. Statist. Assoc.}, \textbf{11}, 581--587.

\item Findley, D. F., Monsell, B. C., Bell, W. R., Otto, M. C., and Chen, B. C. (1998). ^^ ^^ New capabilities and methods of the X-12-ARIMA seasonal-adjustment program,"  \textit{Journal of Business \& Economic Statistics}, \textbf{16}(2), 127--152.

\item Gersch, W., and Kitagawa, G. (1983), ^^ ^^ The prediction of time series with trends and seasonalities,"  \textit{Journal of Business \& Economic Statistics}, \textbf{1}(3), 253--264.

\item Haba, T. and Nagao, H. (2018), ^^ ^^ Estimation of multiple periodic seasonal components in seasonally adjustment models," Abstract of master's thesis, Graduate School of Information Science and Technology, The University of Tokyo (in Japanese).

\item Hillmer S. C. and Tiao G. C. (1982), ^^ ^^ An ARIMA based approach to seasonal adjustment,"  \textit{J. Am. statist. Assoc.}, \textbf{11}, 63--70.

\bibitem{Kitagawa 1987}
Kitagawa, G. (1987). ^^ ^^ Non-Gaussian state-space modeling of nonstationary time series,"
{\it Journal of the American Statistical Association}, 82(400), 1032--1041.

\bibitem{Kitagawa 1989}
Kitagawa, G. (1989). ^^ ^^ Non-Gaussian seasonal adjustment," {\it Computers \& Mathematics with Applications}, 
Vol.18, No.6/7, 503--514.


\bibitem{Kitagawa 1994}
Kitagawa, G. (1994).  ^^ ^^ The two-filter formula for smoothing and an implementation of the Gaussian-sum smoother,"
{\it Annals of the Institute of Statistical Mathematics}, \textbf{46}(4), 605--623.


\bibitem{Kitagawa 1996} 
Kitagawa, G. (1996). ^^ ^^ Monte Carlo filter and smoother for non-Gaussian nonlinear state space models," 
{\it Journal of Computational and Graphical Statistics}, \textbf{5}(1), 1--25.

\bibitem{Kitagawa 2021}
Kitagawa, G. (2021). \textit{Introduction to Time Series Modeling with Applications in R},
Second Edition, Chapman \& Hall CRC Press.

\item Kitagawa, G. and Gersch, W. (1984), ^^ ^^ A smoothness priors-state space modeling of
time series with trend and seasonality," {\it J. Amer. Statist. Assoc.}, {\bf 79}, 378--389.

\item Kitagawa, G. and Gersch, W. (1996), {\it Smoothness Priors Analysis of Time Series},
{\it Lecture Notes in Statistics}, {\bf 116}, Springer, New York.

\item Taylor, J. W. (2010), ^^ ^^ Triple seasonal methods for short-term electricity demand forecasting,"  \textit{European Journal of Operational Research}, \textbf{204}(1), 139--152.

\end{thebibliography}
\end{document}